\documentclass{emulateapj}
\usepackage[modulo]{lineno}
\usepackage[xcolor]{xcolor}
\usepackage{amsmath}

\def\deg{{$^{\circ}$}}

\def\kmsec{\mbox{km~s$^{\rm -1}$}}
\def\logg{\mbox{log~{\it g}}}
\def\msun{\mbox{$M_{\odot}$}}
\def\teff{\mbox{$T_{\rm eff}$}}
\def\vt{\mbox{$v_{\rm t}$}}
\def\cempr{\mbox{CEMP-$r$}}

\def\rtwo{\mbox{$r$-II}}
\def\rpro{\mbox{$r$-process}}
\def\spro{\mbox{$s$-process}}

\def\ncap{\mbox{$n$-capture}}
\def\loggf{\mbox{$\log gf$}}
\def\stara{\mbox{Star~1}}   
\def\starb{\mbox{Star~2}}   
\def\starc{\mbox{Star~3}}   
\def\stard{\mbox{Star~4}}   
\def\ret{\object[NAME Reticulum II]{Ret~2}}
\def\umigal{\object[NAME UMi Galaxy]{UMi}}
\def\booonegal{\object[NAME Bootes Dwarf Spheroidal Galaxy]{Boo~I}}

\def\comgal{\object[NAME Coma Dwarf Galaxy]{Com}}
\def\hergal{\object[NAME Her Dwarf Galaxy]{Her}}

\def\segonegal{\object[NAME Segue 1]{Seg~1}}
\def\segtwogal{\object[NAME Segue 2]{Seg~2}}

\def\umagal{\object[NAME UMa II Galaxy]{UMa~II}}

\def\dragal{\object[NAME Draco Dwarf Spheroidal Galaxy]{Dra}}

\def\cargal{\object[NAME Carina dSph]{Car}}
\def\lmc{\object[NAME LMC]{LMC}}

\def\cs{\object[BPS CS 22892-052]{CS~22892--052}}

\shorttitle{Detailed Abundances in Ret 2}
\shortauthors{Roederer et al.}
\slugcomment{Accepted for publication in the Astronomical Journal}

\begin{document}

\title{
Detailed Chemical Abundances in the $r$-Process-Rich 
Ultra-Faint Dwarf Galaxy
Reticulum 2\footnotemark[$\dagger$]
}

\footnotetext[$\dagger$]{This paper includes data gathered with the 6.5 meter 
Magellan Telescopes located at Las Campanas Observatory, Chile.
}

\author{
Ian U.\ Roederer,\altaffilmark{1,2}
Mario Mateo,\altaffilmark{1}
John I.\ Bailey III,\altaffilmark{1}
Yingyi Song,\altaffilmark{1}
Eric F.\ Bell,\altaffilmark{1}
Jeffrey D.\ Crane,\altaffilmark{3} \\
Sarah Loebman,\altaffilmark{1}
David L.\ Nidever,\altaffilmark{1,4,5} 
Edward W.\ Olszewski,\altaffilmark{5}
Stephen A.\ Shectman,\altaffilmark{3} \\
Ian B.\ Thompson,\altaffilmark{3}
Monica Valluri,\altaffilmark{1}
Matthew G.\ Walker\altaffilmark{6}
}

\altaffiltext{1}{Department of Astronomy, University of Michigan,
1085 S.\ University Ave., Ann Arbor, MI 48109, USA;
\mbox{iur@umich.edu}
}
\altaffiltext{2}{Joint Institute for Nuclear Astrophysics and Center for the
Evolution of the Elements (JINA-CEE), USA
}
\altaffiltext{3}{Carnegie Observatories,
813 Santa Barbara St., Pasadena, CA 91101, USA
}
\altaffiltext{4}{Large Synoptic Survey Telescope, 
950 North Cherry Ave., Tucson, AZ 85719, USA
}
\altaffiltext{5}{Steward Observatory, 
933 North Cherry Ave, Tucson, AZ 85719, USA
}
\altaffiltext{6}{McWilliams Center for Cosmology, 
Department of Physics, Carnegie Mellon University, 5000 Forbes Ave.,
Pittsburgh, PA 15213, USA
}


\addtocounter{footnote}{6}

\begin{abstract}

The ultra-faint dwarf galaxy Reticulum~2 (Ret~2) was
recently discovered in images 
obtained by the Dark Energy Survey.
We have observed the four brightest red giants in
Ret~2 at high spectral resolution using the
Michigan/Magellan Fiber System.
We present detailed abundances for
as many as 20 elements per star, including 12~elements
heavier than the Fe group.
We confirm previous detection of high
levels of $r$-process material
in Ret~2
(mean [Eu/Fe]~$= +$1.69~$\pm$~0.05)
found in three of these stars 
(mean [Fe/H]~$= -$2.88~$\pm$~0.10).
The abundances closely match the
$r$-process pattern found in the
well-studied metal-poor halo star \mbox{CS~22892--052}.
Such $r$-process-enhanced stars 
have not been found in any other ultra-faint dwarf galaxy,
though their existence has been predicted by at least one model.
The fourth star in Ret~2 
([Fe/H]~$= -$3.42~$\pm$~0.20)
contains only trace amounts of Sr
([Sr/Fe]~$= -$1.73~$\pm$~0.43)
and no detectable heavier elements.
One $r$-process enhanced star is 
also enhanced in C
(natal [C/Fe]~$\approx +$1.1).
This is only the third such star known,
which suggests that the nucleosynthesis sites
leading to C and
$r$-process enhancements are decoupled.
The $r$-process-deficient star is enhanced in Mg
([Mg/Fe]~$= +$0.81~$\pm$~0.14),
and the other three stars show normal levels of
$\alpha$-enhancement 
(mean [Mg/Fe]~$= +$0.34~$\pm$~0.03).
The abundances of other $\alpha$ and Fe-group elements 
closely resemble those in ultra-faint dwarf galaxies
and metal-poor halo stars, suggesting 
that the nucleosynthesis
that led to the large $r$-process enhancements
either produced no light elements 
or produced light-element abundance signatures 
indistinguishable from normal supernovae.

\end{abstract}

\keywords{
galaxies:\ dwarf ---
galaxies:\ individual (Reticulum 2) ---
nuclear reactions, nucleosynthesis, abundances ---
stars:\ abundances
}

\section{Introduction}
\label{intro}

Images obtained by the Dark Energy Survey (DES; \citealt{diehl14}) 
have recently revealed 17 new faint Milky Way satellite candidates
toward the general direction of the Magellanic Clouds
\citep{bechtol15,drlicawagner15,koposov15a}.
One of these, Reticulum~2 (\ret),
is the nearest ($\approx$~30~kpc) ultra-faint dwarf galaxy
(UFD; $M_V = -$2.7) 
to the Milky Way in the southern hemisphere.
Reconnaissance spectroscopy of \ret\
\citep{koposov15b,simon15,walker15}
have confirmed \ret\ has low mean metallicity ([Fe/H]~$= -$2.6), 
a significant metallicity dispersion ($\sigma_{\rm [Fe/H]} \approx$~0.5), 
and a high mass-to-light ratio 
($\gtrsim$~500 in solar units).
Though located relatively close to the 
Large Magellanic Cloud (\lmc),
it remains unclear whether \ret\ 
was originally a satellite of the Magellanic system
\citep{koposov15b}.
No H~\textsc{i} has been detected in \ret\ \citep{westmeier15}.
\citet{geringersameth15} report a 
possible gamma-ray signal detection toward \ret\
that may be consistent with dark matter  
annihilation.

\citet{ji15b} presented the first abundances derived from
high-resolution spectroscopy of nine of the brightest stars in \ret.
Their startling finding is that seven of these nine stars
are highly enhanced in material produced by the 
rapid neutron-capture process ($r$~process).
This is remarkable on several levels.

First, no previous studies of dozens of stars in 
nine UFD galaxies
have revealed large enhancements of heavy elements produced by 
\rpro\ nucleosynthesis 
\citep{koch08,feltzing09,frebel10,frebel14,
norris10a,norris10b,simon10,gilmore13,koch13,ishigaki14a,
koch14,roederer14b,francois15,ji15a}.
Typically, elements produced by neutron-capture (\ncap) reactions
are difficult to detect in the UFD galaxies,
and their abundances are among the lowest known in any stars.

Second, stars with such high levels of 
\rpro\ material ([Eu/Fe]~$> +$1 and [Ba/Eu]~$<$~0;
hereafter known as \rtwo\ stars, \citealt{christlieb04})
are relatively rare among the metal-poor stars
in the solar neighborhood
($\sim$~3\%; \citealt{barklem05}).
About 20 \rtwo\ stars are known, and 
all are located in the halo,
with the exception of one star in \umigal\
\citep{shetrone01,sadakane04,aoki07b}
and one in the Milky Way bulge \citep{johnson13}.
Until now,
none of the stars in the \rtwo\ class
have been physically associated with each other,
and finding a galaxy teeming with
such stars is unprecedented.

Third, there have been few
opportunities to test the role that
environment plays in the formation of \rtwo\ stars
and \rpro\ nucleosynthesis more generally.
\citet{ji15b}\
assumed that UFD galaxies like \ret\ formed in minihalos
with $\sim$~10$^{6}$~\msun\ of H (e.g., \citealt{bromm04}).
\citeauthor{ji15b}\ concluded that
the \rpro\ material originated in 
a single neutron-star merger (e.g., \citealt{goriely11,ramirezruiz15})
or any other rare site with high \rpro\ yields
($\sim$~10$^{-4.5}$~\msun\ of Eu per event).
The amount of \rpro\ material required
scales with the assumed H mass of the minihalo.
The dilution of metals into H may not
be uniform across the galaxy, and
all \rpro\ material may not be incorporated into stars,
but this sets a lower limit on the 
amount of \rpro\ material necessary.

Previous attempts to 
identify the astrophysical source(s) of the $r$~process
have focused on 
characterizing the \rpro\ abundance pattern in detail
(e.g., \citealt{sneden96,sneden03,hill02,christlieb04,frebel07,siqueiramello13})
and modeling the physical conditions of
nucleosynthesis to reproduce this pattern
(e.g., \citealt{wanajo03,kratz07,qian07,hayek09}).
Other approaches include
estimating the frequency of \rtwo\ stars in the field
\citep{barklem05},
estimating the frequency of \rtwo\ stars
exhibiting radial velocity variations
that reveal unseen companions
\citep{hansen11,hansen15b},
searching for kinematic properties 
shared by \rpro-enhanced stars
\citep{roederer09a},
searching for subtle abundance patterns among the lighter elements
in \rtwo\ stars
\citep{roederer14e},
probing the extent to which \rpro\ material
can be detected in trace amounts in metal-poor stars
\citep{roederer13,roederer14a},
modeling the light curves associated with \rpro\ events
that might be found in transient surveys
\citep{metzger10,goriely11,kasen15},
and 
searching for short-lived radioactive 
nuclides produced by the $r$~process
in deep-sea crust or sediments
\citep{wallner15,hotokezaka15}.
Others pursue this question by
modeling the chemical evolution of \rpro\ elements
integrated across the Galactic halo
(e.g., \citealt{mathews92,ishimaru99,
argast04,matteucci14,cescutti15}),
within dwarf spheroidal (dSph) galaxies
\citep{tsujimoto14},
or within cosmologically-motivated galaxy 
merger models and simulations
\citep{komiya14,ishimaru15,shen15,vandevoort15}.
The growing preponderance of evidence appears to point towards
neutron star mergers as a dominant source of
\rpro\ material,
but definitive observational evidence is still lacking.

Here, we present a detailed abundance analysis of four stars
in \ret.
We confirm the abundances of Fe, Ba, and Eu
reported by \citet{ji15b},
and we expand the chemical inventory to include 
20 elements from C ($Z =$~6) to Dy ($Z =$~66).
We also present upper limits for six other elements.
We present the new observational material in
Section~\ref{observations},
outline our analysis in Section~\ref{analysis},
describe our results in Section~\ref{results},
and discuss their implications in Section~\ref{discussion}.

\section{Observations}
\label{observations}

There are only four stars brighter than the horizontal branch
that are confirmed members of \ret\ \citep{simon15,walker15}.
Table~\ref{startab} lists the $g$ magnitudes and
$g-r$ colors, 
adopted from the photometric catalogs 
that \citet{koposov15a} generated from public DES images.
The DES names are listed in Table~\ref{startab}.
Throughout this paper,
we refer to these stars by the shorter
names listed in Table~\ref{startab}.

We observed these four stars
using one arm of the Michigan/Magellan Fiber System
(M2FS)
and MSpec double spectrograph \citep{mateo12,bailey12}
mounted on the Nasmyth platform at the 6.5~m Landon Clay Telescope
(Magellan~II) at Las Campanas Observatory, Chile.
We observed four high-probability members of \ret\ and one blank sky position
simultaneously
on 2015 November 14 and 16,
with a total integration time of 6.67~h.
Both observations were taken in dark time.

\begin{figure}
\centering
\includegraphics[angle=0,width=3.1in]{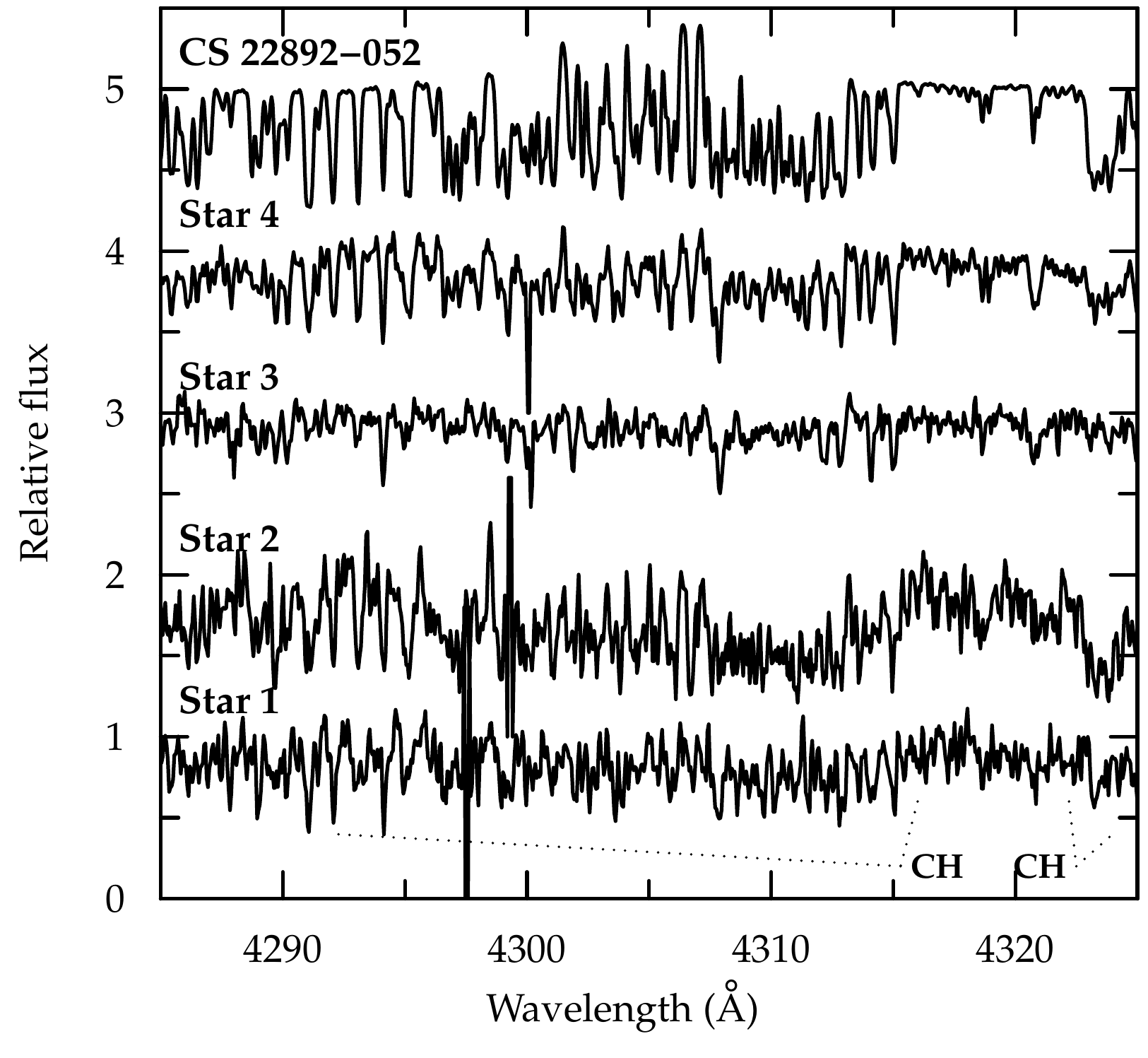} \\
\vspace*{0.05in}
\includegraphics[angle=0,width=3.1in]{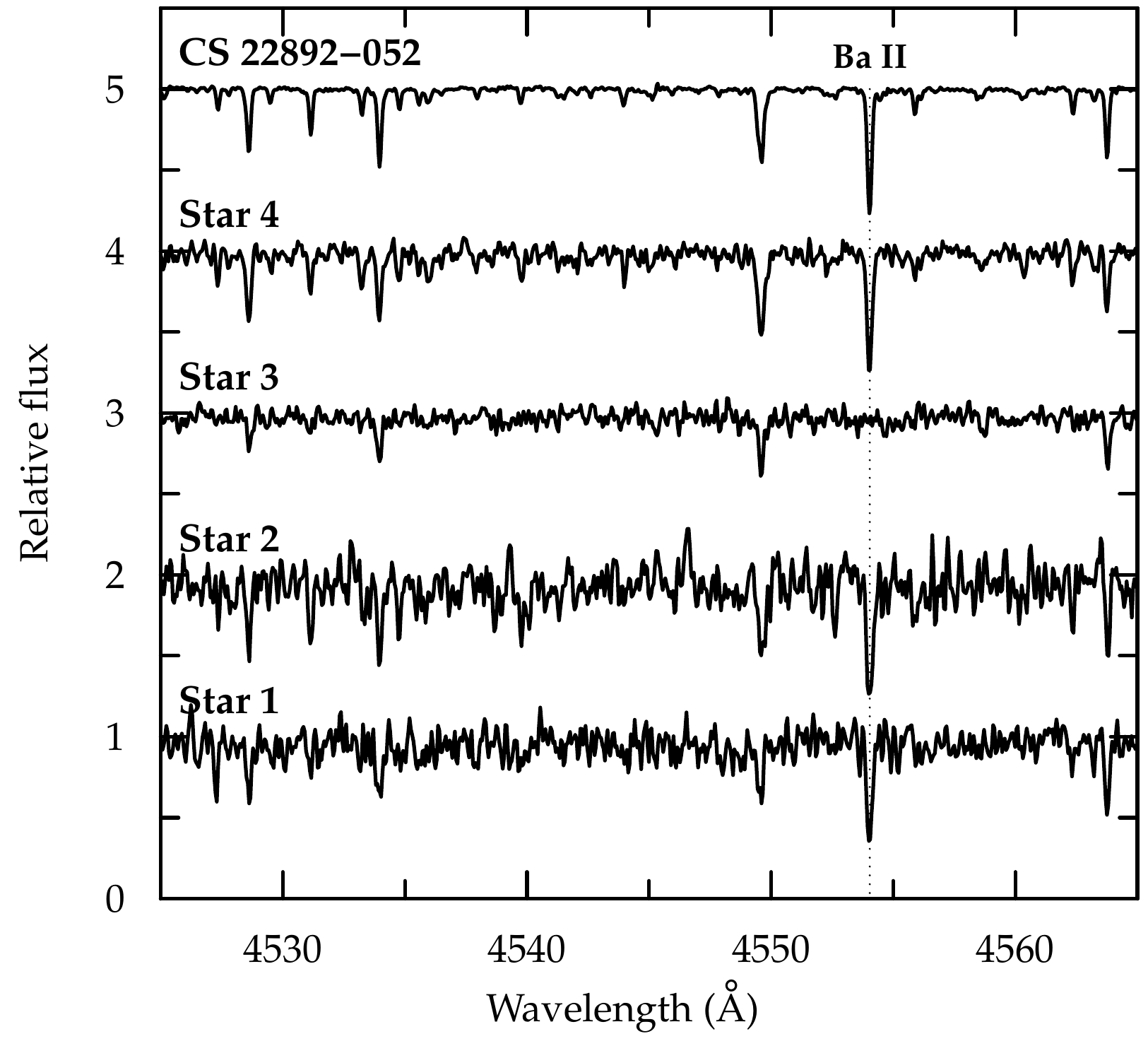} \\
\vspace*{0.05in}
\includegraphics[angle=0,width=3.1in]{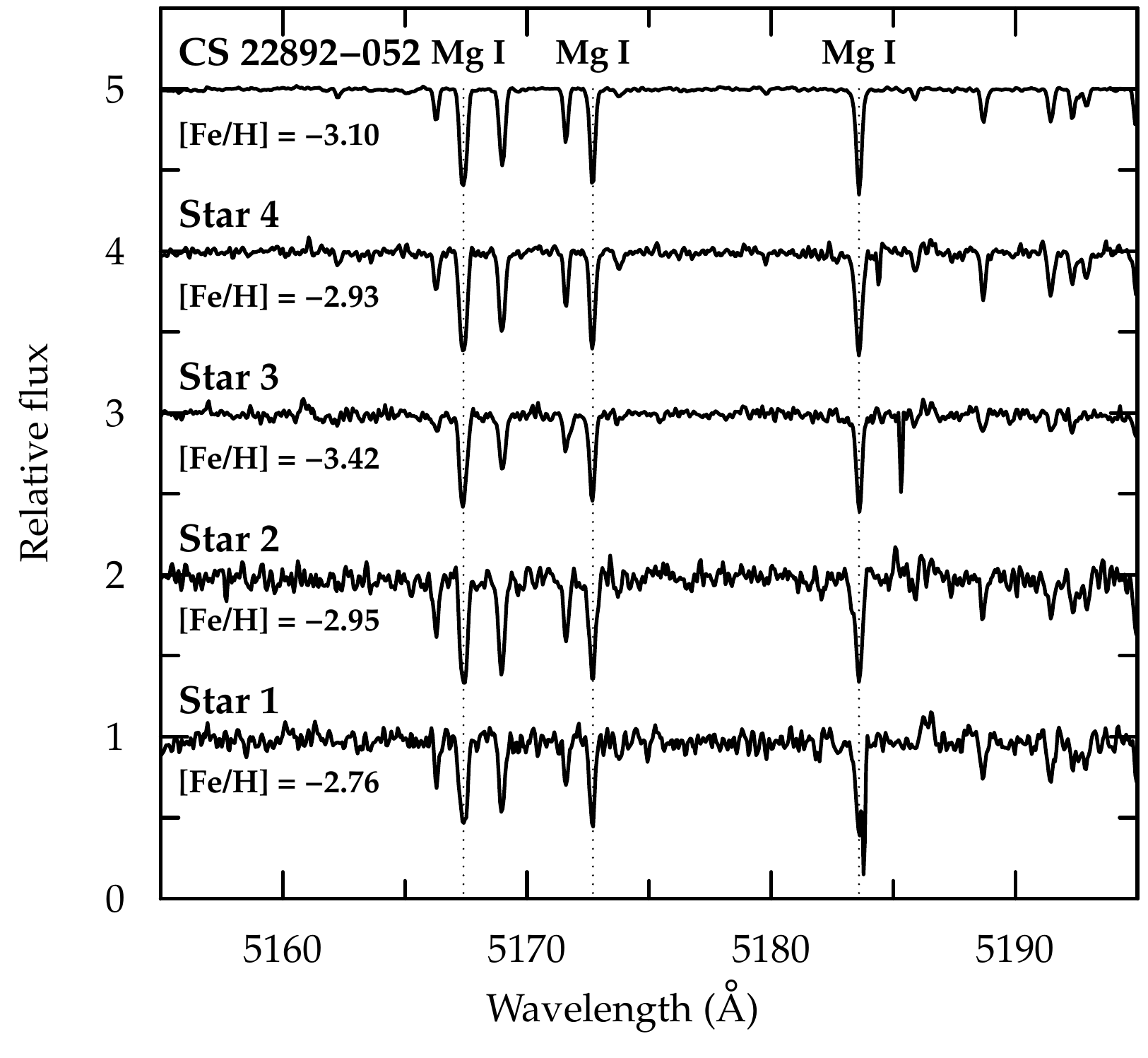} \\
\caption{
\label{m2fsrangeplot}
Selections of the normalized 
M2FS spectra of four stars in \mbox{Ret~2}
and \mbox{CS~22892--052}
around the CH G band (top),
Ba~\textsc{ii} line at 4554~\AA\ (middle),
and Mg~\textsc{i} b triplet (bottom).
The spectra 
have been offset vertically for display purposes.
}
\end{figure}

Our observations were made using the HiRes mode of M2FS
with 95~$\mu$m entrance slits.  
This setup delivered spectral resolving power
$R \equiv \lambda/\Delta\lambda \sim$~30,000,
as measured from isolated Th or Ar emission lines
in the comparison lamp spectra.
We use a new custom order-isolation filter 
to observe orders 66--86,
which covers roughly 4150~$\leq \lambda \leq$~5430~\AA\
for each target.
Figure~\ref{m2fsrangeplot} illustrates sections of the
M2FS spectra of these four stars.

Some data reduction (merging data from 
different CCD chip amplifiers, stacking images,
masking cosmic rays, 
and subtracting scattered light) 
was performed using custom routines.
Standard IRAF routines were used to perform
all other tasks (flatfielding, extraction, wavelength calibration, 
spectra co-addition, velocity shifting, and
continuum normalization).
Sky contamination was found to be negligible,
so no sky subtraction was performed.
Table~\ref{startab} also lists the
signal-to-noise (S/N) ratios
per pixel for the co-added spectra
at several wavelengths.

\begin{deluxetable*}{ccccccccc}
\tablecaption{Basic Stellar Data
\label{startab}}
\tablewidth{0pt}
\tabletypesize{\scriptsize}
\tablehead{
\colhead{Name\tablenotemark{a}} &
\colhead{Name\tablenotemark{b}} &
\colhead{Name\tablenotemark{c}} &
\colhead{$g$} &
\colhead{$g-r$} &
\colhead{$V_{r}$} &
\colhead{S/N} &
\colhead{S/N} &
\colhead{S/N} \\
\colhead{} &
\colhead{} &
\colhead{} &
\colhead{} &
\colhead{} &
\colhead{(\kmsec)} &
\colhead{4300~\AA} &
\colhead{4800~\AA} &
\colhead{5200~\AA} 
}
\startdata
Star 1 & DES J033447.94$-$540525.0 &  Ret2-80 & 17.50 & 0.59 & $+$62.2 & 17 & 26 & 33  \\
Star 2 & DES J033523.85$-$540407.5 &  \nodata & 16.47 & 0.80 & $+$65.5 & 15 & 26 & 35  \\
Star 3 & DES J033531.14$-$540148.2 & Ret2-115 & 17.57 & 0.56 & $+$59.7 & 25 & 40 & 48  \\
Star 4 & DES J033607.75$-$540235.6 & Ret2-178 & 17.39 & 0.58 & $+$62.0 & 26 & 45 & 59  
\enddata
\tablenotetext{a}{This study}
\tablenotetext{b}{\citet{simon15}}
\tablenotetext{c}{\citet{walker15}}
\end{deluxetable*}

\section{Analysis}
\label{analysis}

\subsection{Radial Velocities}
\label{rv}

We measure radial velocities 
using the IRAF \textit{fxcor} task to cross-correlate
the order containing the Mg~\textsc{i} b lines
against a template.
We use the highest S/N spectrum in our sample
(\stard, observation 2)
as the template.
We set the zeropoint of this spectrum by 
measuring the offset between observed and
laboratory wavelengths of several Mg~\textsc{i} and Fe~\textsc{i}
lines.
We adopt laboratory wavelengths from the
Atomic Spectra Database (ASD) of the 
National Institute of Standards and Technology (NIST; \citealt{kramida15}).
The velocity zeropoint of the template is 
accurate to $\sim$~0.2~\kmsec.
We reproduce the radial velocities
of two standard stars 
(\object[CD-43 2527]{CD$-$43\deg2527} and
\object[HD 48381]{HD~48381})
observed on each night 
to within 0.5~\kmsec\ r.m.s.\
\citep{udry99}.
\citet{roederer16} reported 
radial velocity measurement uncertainties
$\sim$~0.7--1.0~\kmsec\
using the 
same M2FS entrance slits
based on repeat observations of stars
with comparable S/N ratios.
We estimate that the uncertainties on 
our radial velocity measurements
do not exceed 1.0~\kmsec.
The two velocity measurements for each star
show no evidence of variations.

We report mean heliocentric radial velocities, $V_{r}$, for each star
in Table~\ref{startab}.
We calculate heliocentric corrections 
using the IRAF \textit{rvcorrect} task.
The velocities in Table~\ref{startab} 
represent an unweighted mean of our two observations.
We find reasonable agreement between our
radial velocities and those measured by
\citet{koposov15b}, \citet{simon15}, and \citet{walker15},
as reported in 
Table~\ref{comparetab}.

\begin{deluxetable}{ccccc}
\tablecaption{Comparison with Previous Studies
\label{comparetab}}
\tablewidth{0pt}
\tabletypesize{\scriptsize}
\tablehead{
\colhead{Quantity} &
\colhead{Study} &
\colhead{$\langle\Delta\rangle$\tablenotemark{a}} &
\colhead{$\sigma$} &
\colhead{N} 
}
\startdata
$V_{r}$ (\kmsec) & \citet{simon15}    & $+$1.2~$\pm$~0.8 & 1.6 & 4 \\
                 & \citet{walker15}   & $-$0.3~$\pm$~0.4 & 0.7 & 3 \\
                 & \citet{koposov15b} & $-$0.3         &\nodata& 2 \\
\teff\ (K) & \citet{ji15b}     & $+$32~$\pm$~47     & 93   & 4 \\
\logg\     & \citet{ji15b}     & $+$0.07~$\pm$~0.15 & 0.29 & 4 \\
~[Fe/H]    & \citet{simon15}   & $-$0.20~$\pm$~0.10 & 0.18 & 3 \\
           & \citet{walker15}  & $-$0.22~$\pm$~0.28 & 0.48 & 3 \\
           & \citet{koposov15b}& $-$0.68           &\nodata& 2 \\
           & \citet{ji15b}     & $+$0.04~$\pm$~0.05 & 0.09 & 4 \\
~[Ba/Fe]   & \citet{ji15b}     & $-$0.11~$\pm$~0.20 & 0.35 & 3 \\
~[Eu/Fe]   & \citet{ji15b}     & $-$0.11~$\pm$~0.07 & 0.11 & 3 
\enddata
\tablenotetext{a}{Differences in the sense of
our study minus previous study}
\end{deluxetable}

\subsection{Equivalent Widths and Atomic Data}
\label{ew}

We measure equivalent widths (EWs) 
from the spectra using a semi-automated 
routine that fits Voigt (or Gaussian) 
line profiles to continuum-normalized spectra.
As discussed in \citet{roederer14c},
the user must visually inspect all lines,
and poor fits can be modified by hand.
Table~\ref{ewtab} lists the EWs measured from our data.
\citet{roederer16} showed that the EWs measured from
M2FS HiRes spectra taken with the same 95~$\mu$m entrance slits
agree to better than 1~m\AA\
with those derived from MIKE spectra,
which were shown by \citet{bedell14} and \citet{roederer14c} 
to agree with
high-resolution spectra taken at several observatories.

\begin{deluxetable*}{ccccccccc}
\tablecaption{Equivalent Widths and Atomic Data
\label{ewtab}}
\tablewidth{0pt}
\tabletypesize{\scriptsize}
\tablehead{
\colhead{Species} &
\colhead{$\lambda$} &
\colhead{E.P.} &
\colhead{\loggf} &
\colhead{Ref.} &
\colhead{EW Star 1} &
\colhead{EW Star 2} &
\colhead{EW Star 3} &
\colhead{EW Star 4} \\
\colhead{} &
\colhead{(\AA)} &
\colhead{(eV)} &
\colhead{} &
\colhead{} &
\colhead{(m\AA)} &
\colhead{(m\AA)} &
\colhead{(m\AA)} &
\colhead{(m\AA)} 
}
\startdata
Mg~\textsc{i} &   4571.10 &  0.00 & $-$5.62  &  1 &    45.2 &    90.8 &    53.0 &    39.2 \\
Mg~\textsc{i} &   4702.99 &  4.33 & $-$0.38  &  1 &    63.9 &    55.7 &    59.5 &    71.1 
\enddata
\tablecomments{The word ``synth'' in the
EW columns indicates that abundances were derived 
by spectral synthesis matching.
The word ``limit'' indicates that
a 3$\sigma$ upper limit was derived from a non-detection. 
\\
The complete version of Table~\ref{ewtab} is
available in the online edition of the Journal. 
A short version is shown here to illustrate its form and content.
}
\tablerefs{(1) \citealt{kramida15};
 (2) \citealt{aldenius07};
 (3) \citealt{lawler89}, using HFS from \citealt{kurucz95};
 (4) \citealt{lawler13};
 (5) \citealt{wood13};
 (6) \citealt{lawler14};
 (7) \citealt{sobeck07};
 (8) \citealt{nilsson06};
 (9) \citealt{booth84};
(10) \citealt{denhartog11} for both \loggf\ value and HFS;
(11) \citealt{ruffoni14};
(12) \citealt{wood14};
(13) \citealt{kramida15}, using HFS from \citealt{kurucz95};
(14) \citealt{roederer12a};
(15) \citealt{biemont11};
(16) \citealt{ljung06};
(17) \citealt{palmeri05};
(18) \citealt{wickliffe94};
(19) \citealt{kramida15}, using HFS/IS from \citealt{mcwilliam98} 
        when available;
(20) \citealt{lawler01a}, using HFS from \citealt{ivans06};
(21) \citealt{lawler09}; 
(22) \citealt{li07}, using HFS from \citealt{sneden09};
(23) \citealt{ivarsson01}, using HFS from \citealt{sneden09};
(24) \citealt{denhartog03}, using HFS/IS from \citealt{roederer08} 
        when available;
(25) \citealt{lawler06}, using HFS/IS from \citealt{roederer08} when available;
(26) \citealt{lawler01b}, using HFS/IS from \citealt{ivans06};
(27) \citealt{denhartog06};
(28) \citealt{lawler01c}, using HFS from \citealt{lawler01d} when available;
(29) \citealt{wickliffe00}
}
\end{deluxetable*}

Table~\ref{ewtab} also includes the 
wavelengths ($\lambda$), 
identifications, excitation potentials (E.P.),
and \loggf\ values for all lines.
We use \loggf\ values from recent laboratory
studies whenever possible,
since these investigations frequently deliver
$\sim$~5\% precision (0.02~dex) or better
(e.g., \citealt{lawler09}).

\subsection{Stellar Parameters and Abundance Analysis}
\label{modelatm}

We base our estimates of the stellar parameters
on measures that can be extracted from the
spectra themselves.
We interpolate model atmospheres from the 
one-dimensional plane-parallel ATLAS9 grid of
$\alpha$-enhanced models \citep{castelli03}.
We compute Fe abundances from Fe~\textsc{i} and Fe~\textsc{ii}
lines using a recent version of the spectrum analysis code
MOOG \citep{sneden73,sobeck11}, which
assumes local thermodynamic equilibrium (LTE) 
for the line-forming layers of the atmosphere.

We derive the effective temperatures (\teff),
log of the surface gravity (\logg),
microturbulence (\vt), and
model metallicity ([M/H])
using an iterative process.
We start from an initial guess 
appropriate for typical metal-poor red giant stars,
but this initial set of parameters has no
substantial impact on the final set.
We estimate \teff\ by requiring no 
correlation between the E.P.\ of Fe~\textsc{i} lines
and the abundances derived from them.
We simultaneously require no correlation between
these abundances and the line strength,
which sets \vt.
Spectroscopic \teff\ values like these
are known to be systematically
cooler by a few percent 
than those derived from color-\teff\ relations
(e.g., \citealt{johnson02}).
Recent measures of the angular diameter of the
nearby metal-poor giant 
\object[HD 122563]{HD~122563} by \citet{creevey12}
favor photometric \teff\ scales,
so we correct our spectroscopic \teff\ values to a
photometric scale using the relation given by \citet{frebel13}.
The uncertainty in \teff\ is dominated by 
the scatter in this calibration.

We next calculate \logg\ from the set of
$Y^{2}$ $\alpha$-enhanced isochrones \citep{demarque04}
using our adjusted \teff\ as the input.
We assume a uniform old age (13~Gyr;
cf., e.g., \citealt{brown12}) for all stars.
We propagate uncertainties from \teff\ into \logg\
using 10$^{4}$ Monte Carlo calculations.
The 13~Gyr [Fe/H]~$= -$2.5 and $-$3.3 isochrones
predict \logg\ different by only 0.06~dex
for red giants,
and an assumed age of 10~Gyr instead of 13~Gyr
only changes the predicted \logg\ by 0.03~dex.
These are insignificant compared to the 
uncertainty introduced by \teff.

We then rederive abundances using the new \logg\ value,
recalculating the \vt\ value and ignoring
the slight correlation between E.P.\ and abundance.
We also set [M/H] equal to the 
Fe abundance derived from Fe~\textsc{ii} lines.
(No Fe~\textsc{ii} lines are measurable in \stara,
so we set [M/H] equal to the abundance derived from Fe~\textsc{i} lines.)
We iterate these steps until the model converges.
Throughout this process
we cull lines ($\sim$~10\% of the lines) 
whose derived abundance deviates from the
mean by more than 2$\sigma$.
The final model atmosphere parameters and their uncertainties
are listed in Table~\ref{modeltab}.
Our \teff\ and \logg\ values 
are in excellent agreement with the values derived by \citet{ji15b},
as shown in Table~\ref{comparetab}.

\begin{deluxetable}{ccccc}
\tablecaption{Stellar Parameters
\label{modeltab}}
\tablewidth{0pt}
\tabletypesize{\scriptsize}
\tablehead{
\colhead{Name} &
\colhead{\teff} &
\colhead{\logg} &
\colhead{\vt} &
\colhead{[M/H]} \\
\colhead{} &
\colhead{(K)} &
\colhead{} &
\colhead{(\kmsec)} &
\colhead{} 
}
\startdata
Star 1 & 5020 (140) & 2.09 (0.38) & 2.00 (0.3) & $-$2.7 (0.2) \\
Star 2 & 4710 (140) & 1.22 (0.40) & 2.85 (0.3) & $-$2.9 (0.2) \\
Star 3 & 4855 (140) & 1.63 (0.39) & 2.40 (0.2) & $-$3.3 (0.2) \\
Star 4 & 4810 (140) & 1.50 (0.39) & 2.15 (0.2) & $-$2.8 (0.2) 
\enddata
\end{deluxetable}

We derive most abundances using an adaptation of the
batch mode capabilities of MOOG.
This method generates theoretical EWs that
are forced to match the measured ones
by adjusting the abundance
of each element.
Lines that are blended or broadened by 
hyperfine splitting (HFS) structure or isotope shifts (IS)
are measured by comparing to a set of 
synthetic spectra generated by MOOG. 
We adopt \rpro\ isotopic fractions for Ba, Nd, Sm, and Eu
\citep{sneden08}.
We derive C abundances by matching synthetic spectra
to the observed
CH $A^{2}\Delta - X^{2}\Pi$ G band
from 4290--4330~\AA,
using a line list provided by B.\ Plez 
(2007, private communication).
We derive 3$\sigma$ upper limits based on non-detections using 
the method described in \citet{roederer14c}.

\section{Results}
\label{results}

Table~\ref{abundtab} lists the abundances derived from each line
in each star.
Tables~\ref{meantab1} and \ref{meantab2}
list the mean abundances and their uncertainties.
For element X, the logarithmic abundance is defined
as the number of atoms of element X per 10$^{12}$ H atoms,
$\log\epsilon$(X)~$\equiv \log_{10}(N_{\rm X}/N_{\rm H}) +$12.0.
For elements X and Y, the logarithmic abundance ratio relative to the
solar ratio is defined as
[X/Y]~$\equiv \log_{10} (N_{\rm X}/N_{\rm Y}) -
\log_{10} (N_{\rm X}/N_{\rm Y})_{\odot}$.
These ratios are constructed 
by referencing abundances derived from
species in the same ionization state
(i.e., neutrals to neutrals and ions to ions).
We adopt the solar reference abundances
given in \citet{asplund09}.

Four sets of uncertainties are listed in Tables~\ref{meantab1}
and \ref{meantab2}.
These are calculated 
as described in \citet{roederer14c},
based on the methodology of \citet{mcwilliam95}.
The statistical uncertainty, $\sigma_{\rm stat}$, 
is given by equation~A17 of \citeauthor{mcwilliam95}\
and includes uncertainties in the EW,
line profile fitting, and \loggf\ values.
The total uncertainty, $\sigma_{\rm tot}$, 
is given by equation~A16 of \citeauthor{mcwilliam95}\
and includes the statistical uncertainty and 
uncertainties in the model atmosphere parameters.
We use the other two uncertainties,
$\sigma_{\rm I}$ and $\sigma_{\rm II}$,
when constructing abundance ratios among different elements.
We add $\sigma_{\rm I}$ for element X 
in quadrature with $\sigma_{\rm stat}$ for
element Y when computing the ratio [X/Y] when Y is 
derived from lines of the neutral species.
Similarly, we add
$\sigma_{\rm II}$ for element X
in quadrature with $\sigma_{\rm stat}$ for
element Y when Y
is derived from lines of the ionized species.

\begin{deluxetable}{cccccc}
\tablecaption{Abundances Derived from Individual Lines
\label{abundtab}}
\tablewidth{0pt}
\tabletypesize{\scriptsize}
\tablehead{
\colhead{Star} &
\colhead{Species} &
\colhead{$\lambda$ (\AA)} &
\colhead{$\log\epsilon$} &
\colhead{$\sigma\log\epsilon$} 
}
\startdata
 Star 1               &     Mg~\textsc{i}  & 4571.10 &    $+$5.28 &    0.20 \\
 Star 1               &     Mg~\textsc{i}  & 4702.99 &    $+$5.14 &    0.26 
\enddata
\tablecomments{The complete version of Table~\ref{abundtab} is
available in the online edition of the Journal. 
A short version is shown here to illustrate its form and content.
}
\end{deluxetable}

\begin{deluxetable*}{cccccccccccccccc}
\tablecaption{Mean Abundances I
\label{meantab1}}
\tablewidth{0pt}
\tabletypesize{\scriptsize}
\tablehead{
\colhead{} &
\multicolumn{7}{c}{Star 1} &
\colhead{} &
\multicolumn{7}{c}{Star 2} \\
\cline{2-8} \cline{10-16}
\colhead{Species} &
\colhead{N} &
\colhead{$\log\epsilon$} &
\colhead{[X/Fe]\tablenotemark{a}} &
\colhead{$\sigma_{\rm stat}$} &
\colhead{$\sigma_{\rm tot}$} &
\colhead{$\sigma_{\rm I}$} &
\colhead{$\sigma_{\rm II}$} &
\colhead{} &
\colhead{N} &
\colhead{$\log\epsilon$} &
\colhead{[X/Fe]\tablenotemark{a}} &
\colhead{$\sigma_{\rm stat}$} &
\colhead{$\sigma_{\rm tot}$} &
\colhead{$\sigma_{\rm I}$} &
\colhead{$\sigma_{\rm II}$} 
}
\startdata
 Fe~\textsc{i}  &  21 &  $+$4.74 & $-$2.76 &  0.09 &  0.19 &  0.00 &  0.00  & &  30 &  $+$4.55 & $-$2.95 &  0.11 &  0.21 &  0.00 &  0.00  \\
 Fe~\textsc{ii} &   0 &  \nodata &  \nodata&\nodata&\nodata&\nodata&\nodata & &   4 &  $+$4.55 & $-$2.95 &  0.14 &  0.20 &  0.00 &  0.00  \\
 C~(CH)         &   1 &  $+$5.92 & $+$0.25 &  0.25 &  0.31 &  0.31 &  0.31  & &   1 &  $+$6.08 & $+$0.60 &  0.25 &  0.32 &  0.32 &  0.32  \\
 Mg~\textsc{i}  &   3 &  $+$5.17 & $+$0.33 &  0.08 &  0.20 &  0.12 &  0.12  & &   4 &  $+$5.03 & $+$0.38 &  0.09 &  0.24 &  0.15 &  0.24  \\
 Ca~\textsc{i}  &   0 &  \nodata &  \nodata&\nodata&\nodata&\nodata&\nodata & &   0 &  \nodata &  \nodata&\nodata&\nodata&\nodata&\nodata \\
 Sc~\textsc{ii} &   0 &  \nodata &  \nodata&\nodata&\nodata&\nodata&\nodata & &   1 & $<+$0.55 &$<+$0.35 &  0.21 &  0.26 &  0.31 &  0.26  \\
 Ti~\textsc{i}  &   4 &  $+$2.85 & $+$0.66 &  0.08 &  0.19 &  0.12 &  0.12  & &   4 &  $+$2.57 & $+$0.56 &  0.11 &  0.21 &  0.15 &  0.25  \\
 Ti~\textsc{ii} &   3 &  $+$2.31 & $+$0.12 &  0.11 &  0.19 &  0.22 &  0.12  & &   3 &  $+$2.38 & $+$0.39 &  0.10 &  0.17 &  0.23 &  0.17  \\
 V~\textsc{i}   &   0 &  \nodata &  \nodata&\nodata&\nodata&\nodata&\nodata & &   0 &  \nodata &  \nodata&\nodata&\nodata&\nodata&\nodata \\
 Cr~\textsc{i}  &   2 &  $+$2.64 & $-$0.23 &  0.10 &  0.19 &  0.13 &  0.13  & &   3 & $<+$2.96 &$<+$0.27 &\nodata&\nodata&\nodata&\nodata \\
 Cr~\textsc{ii} &   0 &  \nodata &  \nodata&\nodata&\nodata&\nodata&\nodata & &   0 &  \nodata &  \nodata&\nodata&\nodata&\nodata&\nodata \\
 Mn~\textsc{i}  &   0 &  \nodata &  \nodata&\nodata&\nodata&\nodata&\nodata & &   0 &  \nodata &  \nodata&\nodata&\nodata&\nodata&\nodata \\
 Ni~\textsc{i}  &   0 &  \nodata &  \nodata&\nodata&\nodata&\nodata&\nodata & &   1 & $<+$4.57 &$<+$1.30 &\nodata&\nodata&\nodata&\nodata \\
 Zn~\textsc{i}  &   0 &  \nodata &  \nodata&\nodata&\nodata&\nodata&\nodata & &   0 &  \nodata &  \nodata&\nodata&\nodata&\nodata&\nodata \\
 Sr~\textsc{ii} &   1 &  $+$0.70 & $+$0.59 &  0.48 &  0.51 &  0.51 &  0.47  & &   1 &  $+$0.46 & $+$0.54 &  0.23 &  0.26 &  0.27 &  0.30  \\
 Y~\textsc{ii}  &   3 &  $-$0.13 & $+$0.42 &  0.16 &  0.21 &  0.26 &  0.17  & &   5 &  $-$0.34 & $+$0.41 &  0.19 &  0.24 &  0.29 &  0.23  \\
 Zr~\textsc{ii} &   1 &  $+$0.69 & $+$0.87 &  0.37 &  0.40 &  0.42 &  0.38  & &   2 & $<+$0.65 &$<+$1.02 &\nodata&\nodata&\nodata&\nodata \\
 Tc~\textsc{i}  &   0 &  \nodata &  \nodata&\nodata&\nodata&\nodata&\nodata & &   0 &  \nodata &  \nodata&\nodata&\nodata&\nodata&\nodata \\
 Ru~\textsc{i}  &   1 & $<+$2.35 &$<+$3.36 &\nodata&\nodata&\nodata&\nodata & &   1 & $<+$1.93 &$<+$3.13 &\nodata&\nodata&\nodata&\nodata \\
 Ba~\textsc{ii} &   1 &  $+$0.22 & $+$0.80 &  0.17 &  0.24 &  0.25 &  0.17  & &   1 &  $+$0.32 & $+$1.09 &  0.21 &  0.24 &  0.25 &  0.28  \\
 La~\textsc{ii} &   1 &  $-$0.24 & $+$1.42 &  0.17 &  0.22 &  0.26 &  0.18  & &   2 &  $-$0.49 & $+$1.36 &  0.20 &  0.25 &  0.30 &  0.25  \\
 Ce~\textsc{ii} &   1 &  $+$0.09 & $+$1.27 &  0.12 &  0.19 &  0.24 &  0.14  & &   5 &  $-$0.11 & $+$1.26 &  0.19 &  0.24 &  0.29 &  0.23  \\
 Pr~\textsc{ii} &   1 &  $-$0.27 & $+$1.77 &  0.41 &  0.43 &  0.46 &  0.41  & &   2 &  $-$0.66 & $+$1.57 &  0.26 &  0.30 &  0.35 &  0.30  \\
 Nd~\textsc{ii} &   6 &  $+$0.35 & $+$1.69 &  0.14 &  0.19 &  0.24 &  0.15  & &   7 &  $-$0.07 & $+$1.46 &  0.17 &  0.23 &  0.28 &  0.22  \\
 Sm~\textsc{ii} &   2 &  $+$0.11 & $+$1.91 &  0.15 &  0.20 &  0.25 &  0.16  & &   3 &  $-$0.17 & $+$1.82 &  0.18 &  0.24 &  0.28 &  0.23  \\
 Eu~\textsc{ii} &   2 &  $-$0.49 & $+$1.75 &  0.27 &  0.32 &  0.33 &  0.27  & &   2 &  $-$0.75 & $+$1.68 &  0.14 &  0.18 &  0.23 &  0.20  \\
 Gd~\textsc{ii} &   2 & $<+$0.64 &$<+$2.33 &\nodata&\nodata&\nodata&\nodata & &   2 & $<+$0.37 &$<+$2.25 &\nodata&\nodata&\nodata&\nodata \\
 Tb~\textsc{ii} &   0 &  \nodata &  \nodata&\nodata&\nodata&\nodata&\nodata & &   1 & $<-$0.78 &$<+$1.87 &\nodata&\nodata&\nodata&\nodata \\
 Dy~\textsc{ii} &   1 & $<+$0.52 &$<+$2.18 &\nodata&\nodata&\nodata&\nodata & &   1 &  $+$0.09 & $+$1.94 &  0.32 &  0.40 &  0.40 &  0.39  
\enddata
\tablenotetext{a}{[Fe/H] given for Fe~\textsc{i} and Fe~\textsc{ii}}
\end{deluxetable*}

\begin{deluxetable*}{cccccccccccccccc}
\tablecaption{Mean Abundances II
\label{meantab2}}
\tablewidth{0pt}
\tabletypesize{\scriptsize}
\tablehead{
\colhead{} &
\multicolumn{7}{c}{Star 3} &
\colhead{} &
\multicolumn{7}{c}{Star 4} \\
\cline{2-8} \cline{10-16}
\colhead{Species} &
\colhead{N} &
\colhead{$\log\epsilon$} &
\colhead{[X/Fe]\tablenotemark{a}} &
\colhead{$\sigma_{\rm stat}$} &
\colhead{$\sigma_{\rm tot}$} &
\colhead{$\sigma_{\rm I}$} &
\colhead{$\sigma_{\rm II}$} &
\colhead{} &
\colhead{N} &
\colhead{$\log\epsilon$} &
\colhead{[X/Fe]\tablenotemark{a}} &
\colhead{$\sigma_{\rm stat}$} &
\colhead{$\sigma_{\rm tot}$} &
\colhead{$\sigma_{\rm I}$} &
\colhead{$\sigma_{\rm II}$} 
}
\startdata
 Fe~\textsc{i}  &  28 &  $+$4.08 & $-$3.42 &  0.11 &  0.20 &  0.00 &  0.00  & &  51 &  $+$4.57 & $-$2.93 &  0.06 &  0.18 &  0.00 &  0.00  \\
 Fe~\textsc{ii} &   2 &  $+$4.24 & $-$3.26 &  0.10 &  0.17 &  0.00 &  0.00  & &   6 &  $+$4.71 & $-$2.79 &  0.07 &  0.15 &  0.00 &  0.00  \\
 C~(CH)         &   1 &  $+$5.14 & $+$0.13 &  0.15 &  0.25 &  0.19 & 0.19   & &   1 &  $+$5.76 & $+$0.26 &  0.15 &  0.25 &  0.19 &  0.19  \\
 Mg~\textsc{i}  &   4 &  $+$4.99 & $+$0.81 &  0.08 &  0.21 &  0.14 &  0.23  & &   4 &  $+$5.01 & $+$0.34 &  0.05 &  0.19 &  0.09 &  0.20  \\
 Ca~\textsc{i}  &   1 &  $+$3.08 & $+$0.16 &  0.28 &  0.32 &  0.30 &  0.34  & &   2 &  $+$3.76 & $+$0.36 &  0.13 &  0.21 &  0.14 &  0.23  \\
 Sc~\textsc{ii} &   1 & $<+$0.79 &$<+$0.90 &\nodata&\nodata&\nodata&\nodata & &   1 & $<+$0.70 &$<+$0.34 &\nodata&\nodata&\nodata&\nodata \\
 Ti~\textsc{i}  &   3 &  $+$1.88 & $+$0.35 &  0.13 &  0.21 &  0.17 &  0.24  & &  12 &  $+$2.51 & $+$0.50 &  0.05 &  0.17 &  0.08 &  0.20  \\
 Ti~\textsc{ii} &   4 &  $+$1.94 & $+$0.25 &  0.14 &  0.19 &  0.24 &  0.18  & &   9 &  $+$2.55 & $+$0.39 &  0.06 &  0.15 &  0.20 &  0.09  \\
 V~\textsc{i}   &   6 & $<+$1.21 &$<+$0.70 &\nodata&\nodata&\nodata&\nodata & &   6 & $<+$1.24 &$<+$0.24 &\nodata&\nodata&\nodata&\nodata \\
 Cr~\textsc{i}  &   1 & $<+$2.23 &$<+$0.01 &\nodata&\nodata&\nodata&\nodata & &   5 &  $+$2.49 & $-$0.22 &  0.06 &  0.18 &  0.08 &  0.20  \\
 Cr~\textsc{ii} &   1 & $<+$3.35 &$<+$0.97 &\nodata&\nodata&\nodata&\nodata & &   3 &  $+$3.50 & $+$0.65 &  0.06 &  0.15 &  0.20 &  0.09  \\
 Mn~\textsc{i}  &   0 &  \nodata &  \nodata&\nodata&\nodata&\nodata&\nodata & &   2 & $<+$2.38 &$<-$0.12 &\nodata&\nodata&\nodata&\nodata \\
 Ni~\textsc{i}  &   1 & $<+$3.95 &$<+$1.15 &\nodata&\nodata&\nodata&\nodata & &   1 &  $+$3.50 & $+$0.21 &  0.09 &  0.19 &  0.11 &  0.21  \\
 Zn~\textsc{i}  &   3 & $<+$2.14 &$<+$1.00 &\nodata&\nodata&\nodata&\nodata & &   1 &  $+$1.96 & $+$0.33 &  0.13 &  0.21 &  0.14 &  0.23  \\
 Sr~\textsc{ii} &   1 &  $-$2.12 & $-$1.73 &  0.42 &  0.44 &  0.47 &  0.43  & &   1 &  $+$0.24 & $+$0.16 &  0.25 &  0.29 &  0.29 &  0.30  \\
 Y~\textsc{ii}  &   9 & $<-$1.00 &$<+$0.05 &\nodata&\nodata&\nodata&\nodata & &   7 &  $-$0.29 & $+$0.29 &  0.08 &  0.16 &  0.21 &  0.11  \\
 Zr~\textsc{ii} &   2 & $<+$1.22 &$<+$1.90 &\nodata&\nodata&\nodata&\nodata & &   3 &  $+$0.36 & $+$0.57 &  0.19 &  0.23 &  0.26 &  0.21  \\
 Tc~\textsc{i}  &   1 & $<+$0.56 &$<+$3.98 &\nodata&\nodata&\nodata&\nodata & &   1 & $<+$0.73 &$<+$3.66 &\nodata&\nodata&\nodata&\nodata \\
 Ru~\textsc{i}  &   1 & $<+$1.81 &$<+$3.48 &\nodata&\nodata&\nodata&\nodata & &   1 & $<+$1.77 &$<+$2.95 &\nodata&\nodata&\nodata&\nodata \\
 Ba~\textsc{ii} &   1 & $<-$2.59 &$<-$1.51 &\nodata&\nodata&\nodata&\nodata & &   1 &  $+$0.06 & $+$0.67 &  0.09 &  0.16 &  0.17 &  0.18  \\
 La~\textsc{ii} &   8 & $<-$1.08 &$<+$1.08 &\nodata&\nodata&\nodata&\nodata & &   5 &  $-$0.44 & $+$1.26 &  0.14 &  0.20 &  0.24 &  0.16  \\
 Ce~\textsc{ii} &   6 & $<-$0.21 &$<+$1.47 &\nodata&\nodata&\nodata&\nodata & &   5 &  $-$0.13 & $+$1.08 &  0.09 &  0.17 &  0.21 &  0.11  \\
 Pr~\textsc{ii} &   4 & $<-$0.97 &$<+$1.57 &\nodata&\nodata&\nodata&\nodata & &   3 &  $-$0.71 & $+$1.36 &  0.16 &  0.21 &  0.25 &  0.17  \\
 Nd~\textsc{ii} &  16 & $<-$0.56 &$<+$1.28 &\nodata&\nodata&\nodata&\nodata & &  18 &  $-$0.07 & $+$1.30 &  0.09 &  0.16 &  0.21 &  0.11  \\
 Sm~\textsc{ii} &  10 & $<-$1.04 &$<+$1.26 &\nodata&\nodata&\nodata&\nodata & &  14 &  $-$0.27 & $+$1.56 &  0.09 &  0.17 &  0.21 &  0.11  \\
 Eu~\textsc{ii} &   1 & $<-$1.66 &$<+$1.08 &\nodata&\nodata&\nodata&\nodata & &   2 &  $-$0.63 & $+$1.64 &  0.06 &  0.13 &  0.14 &  0.14  \\
 Gd~\textsc{ii} &   3 & $<-$0.56 &$<+$1.63 &\nodata&\nodata&\nodata&\nodata & &   1 &  $-$0.26 & $+$1.46 &  0.18 &  0.23 &  0.27 &  0.20  \\
 Tb~\textsc{ii} &   1 & $<-$0.52 &$<+$2.44 &\nodata&\nodata&\nodata&\nodata & &   0 & $<-$0.39 &$<+$2.10 &\nodata&\nodata&\nodata&\nodata \\
 Dy~\textsc{ii} &   1 & $<-$0.24 &$<+$1.92 &\nodata&\nodata&\nodata&\nodata & &   1 &  $+$0.19 & $+$1.88 &  0.12 &  0.20 &  0.22 &  0.18 
\enddata
\tablenotetext{a}{[Fe/H] given for Fe~\textsc{i} and Fe~\textsc{ii}}
\end{deluxetable*}

Our derived metallicities are in reasonable agreement with
those derived by \citet{ji15b}, 
\citet{simon15}, and \citet{walker15}.
The mean metallicity difference calculated from
two stars in common with \citet{koposov15b}
is $-$0.68 dex.
Table~\ref{comparetab} also lists these comparisons.
We are encouraged by the general agreement
in light of the
different methods, spectral coverage, and S/N ratios that 
are used to derive [Fe/H].
Table~\ref{comparetab} also compares our derived [Ba/Fe] and
[Eu/Fe] ratios with those of \citeauthor{ji15b} 
These values are in good agreement.

\subsection{Neutron-Capture Elements}
\label{heavy}

\begin{figure*}
\begin{center}
\includegraphics[angle=0,width=5.90in]{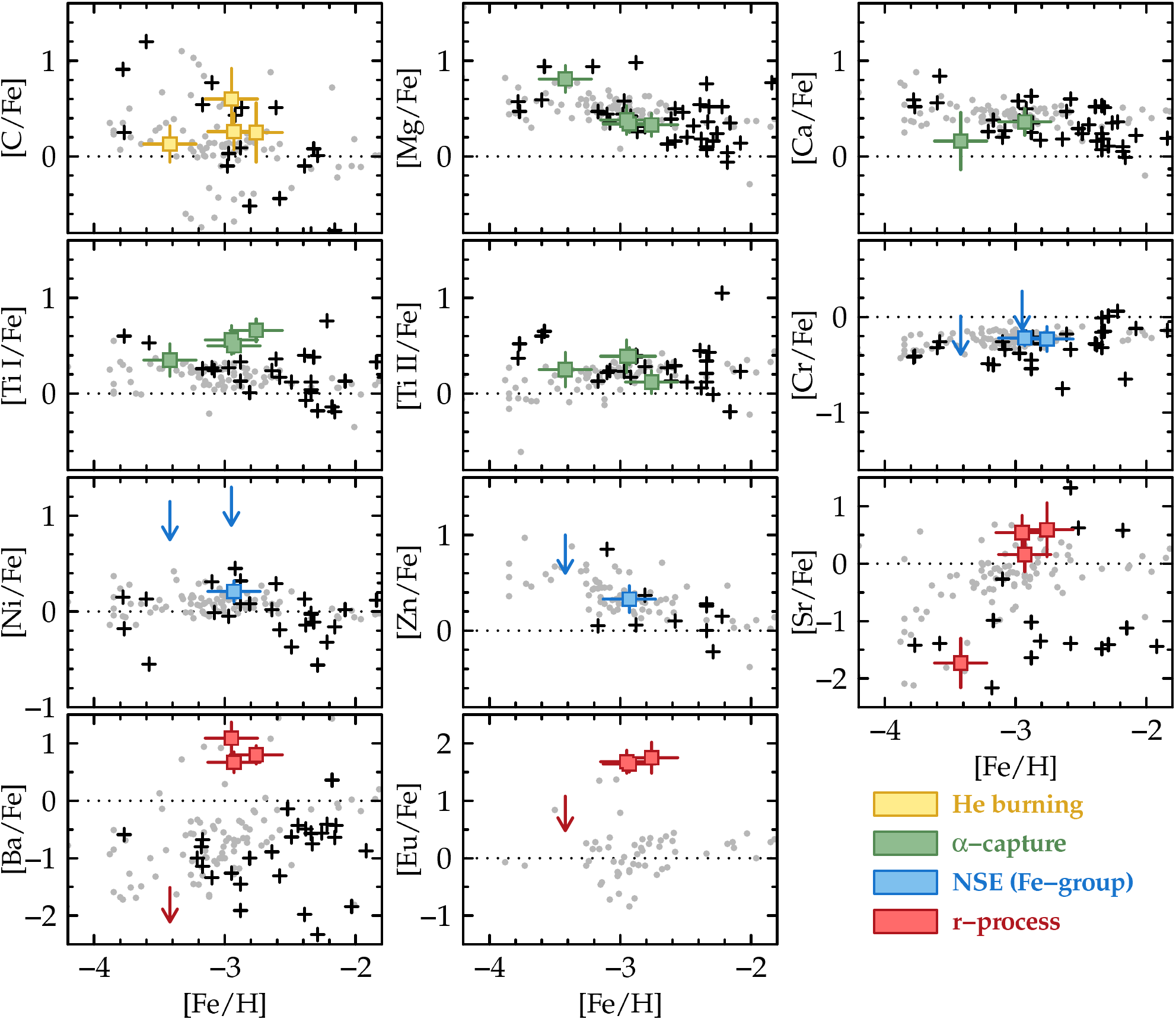}
\end{center}
\caption{
\label{multiplot}
Comparison of abundance ratios in \mbox{Ret~2}
(colored points)
with stars in other UFD galaxies (black crosses)
and halo giants (gray dots).
The UFD sample includes data from
\mbox{Boo~I} \citep{feltzing09,norris10a,gilmore13,ishigaki14a},
\mbox{Boo~II} \citep{koch14,ji15a},
\mbox{CVn~II} \citep{francois15},
\mbox{Com} \citep{frebel10},
\mbox{Her} \citep{koch08,koch13,francois15},
\mbox{Leo~IV} \citep{simon10},
\mbox{Seg~1} \citep{norris10b,frebel14},
\mbox{Seg~2} \citep{roederer14b},
and
\mbox{UMa~II} \citep{frebel10}.
The halo sample includes only the giants
from \citet{roederer14c}.
Upper limits are omitted from the comparison samples
for clarity.
The colors of the \mbox{Ret~2} data points indicate the
dominant nucleosynthetic origins of each element:\
yellow, He-burning; 
green, C and O burning and
$\alpha$-capture in hydrostatic or explosive nucleosynthesis;
blue, Fe-group elements formed in 
nuclear statistical equilibrium (NSE) during explosive nucleosynthesis;
red, \rpro\ nucleosynthesis.
The dotted lines mark the solar ratios.
Note the expanded scale on the vertical axes
for the [Sr/Fe], [Ba/Fe], and [Eu/Fe] panels.
}
\end{figure*}

\begin{figure*}
\begin{center}
\includegraphics[angle=0,width=7.0in]{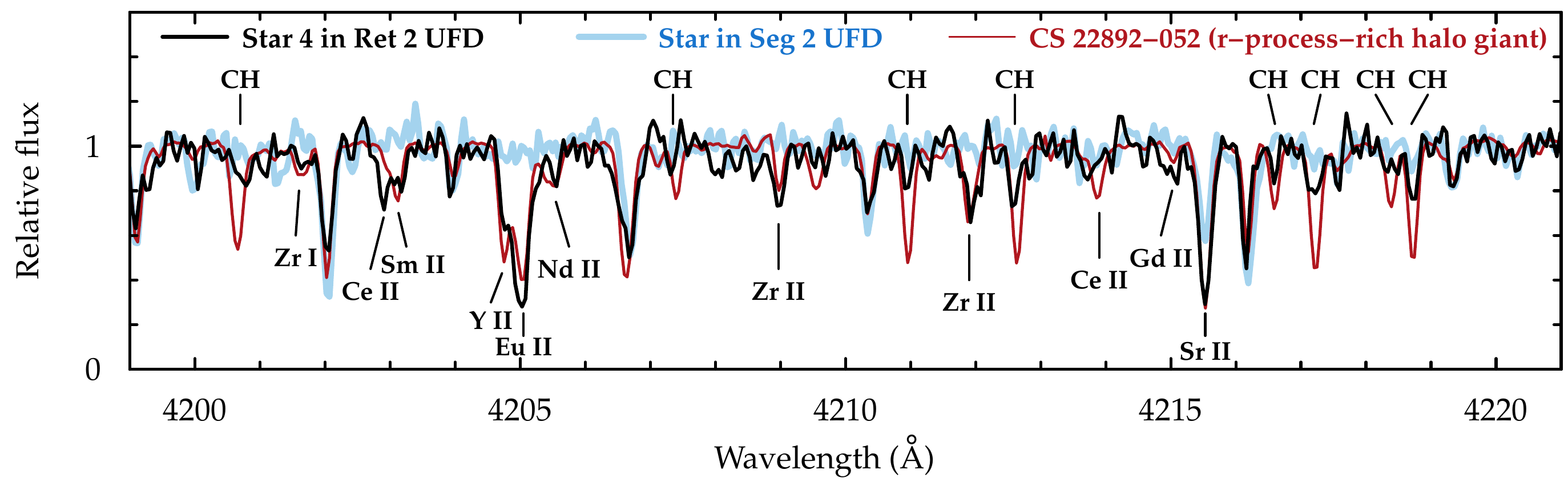}
\end{center}
\caption{
\label{compareplot}
Spectra of \stard\ in \mbox{Ret~2} (black line),
star \mbox{SDSS~J021933.13$+$200830.2} in \mbox{Seg~2} (bold blue line),
and the field giant \mbox{CS~22892--052} (thin red line).
The \mbox{SDSS~J021933.13$+$200830.2}
and \mbox{CS~22892--052} spectra were taken using the 
$R \sim$~41,000 setting of the 
Magellan Inamori Kyocera Echelle (MIKE) spectrograph
(see \citealt{roederer14b} and \citealt{roederer14c}\ for details).
All stars have similar parameters and metallicities,
but their C and \ncap\ abundances differ dramatically.
Lines of \ncap\ species are marked below the spectra,
lines of CH are marked above,
and unmarked lines arise from 
$\alpha$- or Fe-group elements.
 }
\end{figure*}

Figure~\ref{multiplot}
compares the [Sr/Fe], [Ba/Fe], and [Eu/Fe] ratios
in the \ret\ stars with
those in other UFD galaxies and
halo stars in the solar neighborhood.
The Sr~\textsc{ii} line at 4215~\AA\
is detectable (Figure~\ref{compareplot})
in the most metal-poor star in \ret, \starc, 
but no other transitions of any \ncap\ element
are detectable in our spectrum of this star.
The upper limit derived from the Ba~\textsc{ii} line
at 4554~\AA\
is among the lowest for any stars with [Fe/H]~$< -$3
in the UFD galaxies.

The \ncap\ abundances 
observed in the other three stars in \ret\
are unprecedented 
among the UFD galaxies.
All three show extremely enhanced ratios of
[Ba/Fe], [Eu/Fe], and other heavy elements.
Figure~\ref{compareplot} compares a
piece of the spectrum of \stard\
(\teff~$=$~4810~K, [Fe/H]~$= -$2.93, [Eu/Fe]~$= +$1.64)
with two other metal-poor red giants.
One star is in the UFD galaxy
\segtwogal\
(\teff~$=$~4566~K, [Fe/H]~$= -$2.96, [Eu/Fe]~$< -$0.30; 
\citealt{roederer14b}).
That study detected a total of seven lines of \ncap\ elements 
in the star in \segtwogal:\
2~lines of Sr~\textsc{ii} and 5~lines of Ba~\textsc{ii}.
In contrast, numerous lines of \ncap\ elements
are detectable in \stard\ in \ret.
The spectrum of this star closely resembles that
of the metal-poor \rpro-enhanced field giant 
\cs\
(\teff~$=$~4800~K, [Fe/H]~$= -$3.10, [Eu/Fe]~$= +$1.64;
\citealt{sneden03}).
The only substantial differences between the spectra of
\stard\ and \cs\ are the CH features,
since \cs\ is C-enhanced ([C/Fe]~$= +$0.88,
\citeauthor{sneden03}).
These three stars in \ret\ have a mean metallicity
([Fe/H]$= -$2.88~$\pm$~0.10)
in the same range as most \rtwo\ stars
found in the halo,
roughly $-$3.2~$\lesssim$~[Fe/H]~$\lesssim -$2.6
(e.g., \citealt{barklem05,roederer14e}).

\begin{figure*}
\begin{center}
\includegraphics[angle=0,width=2.60in]{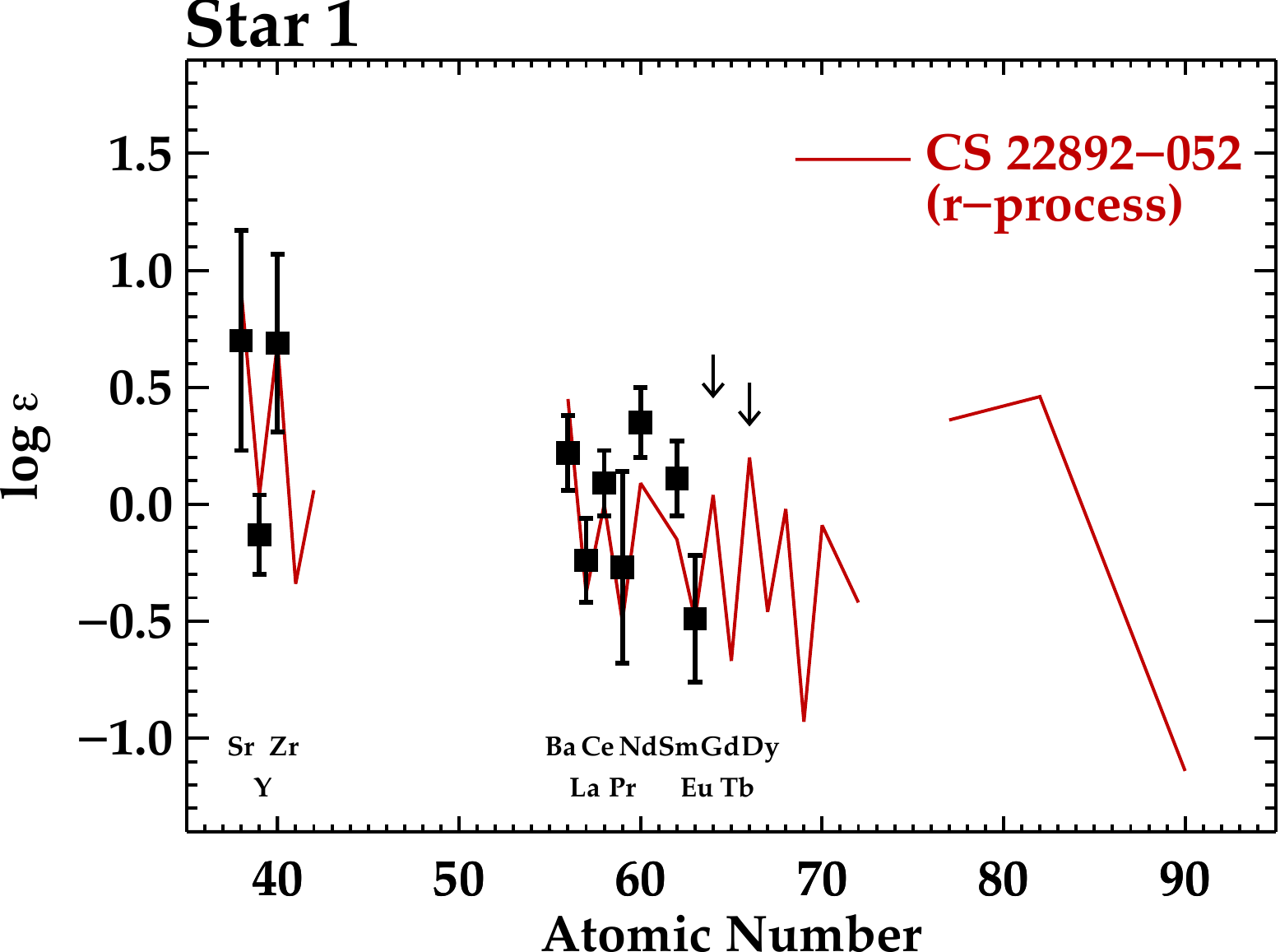}
\hspace*{0.1in}
\includegraphics[angle=0,width=2.60in]{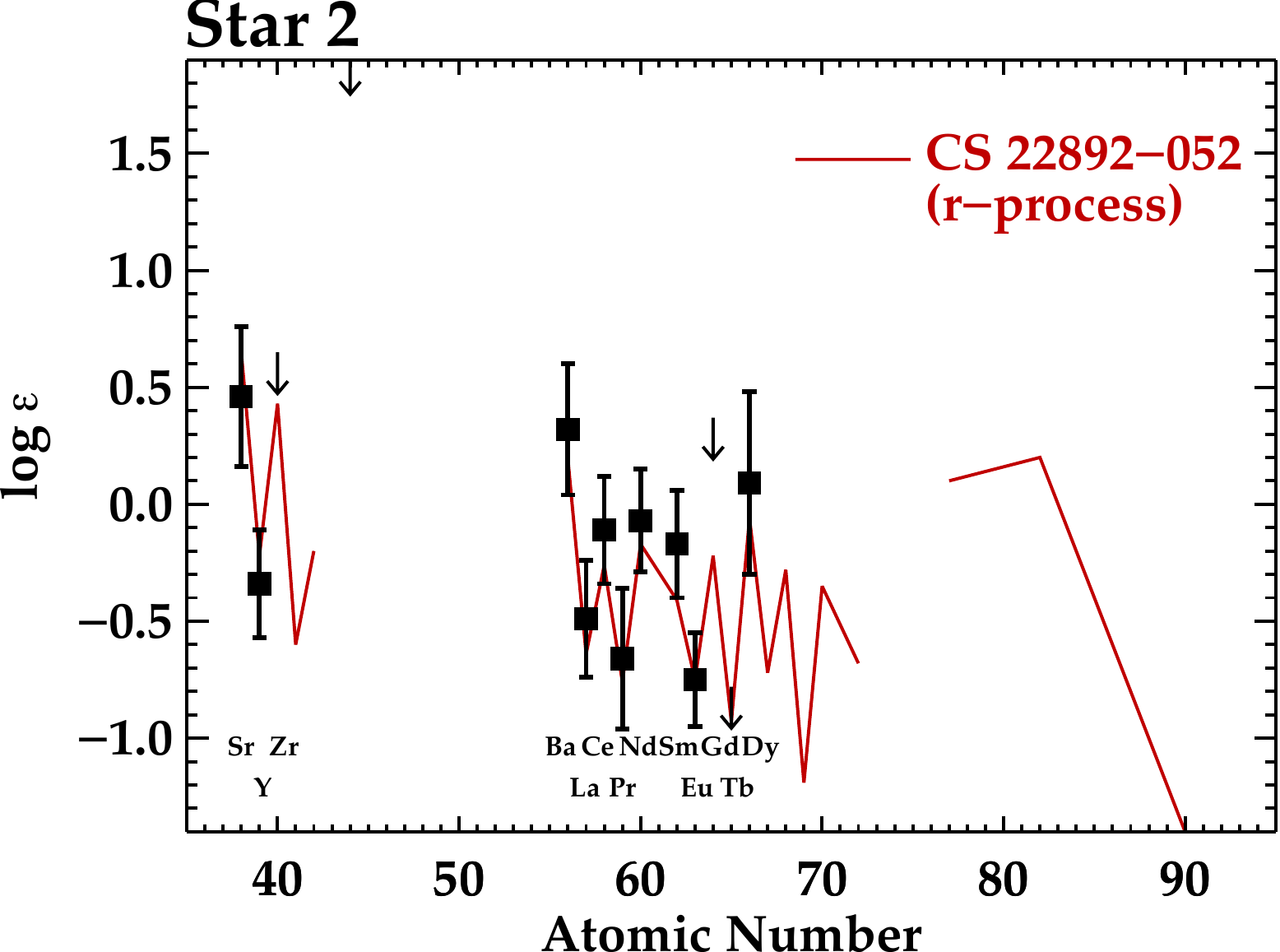} \\
\vspace*{0.1in}
\includegraphics[angle=0,width=2.60in]{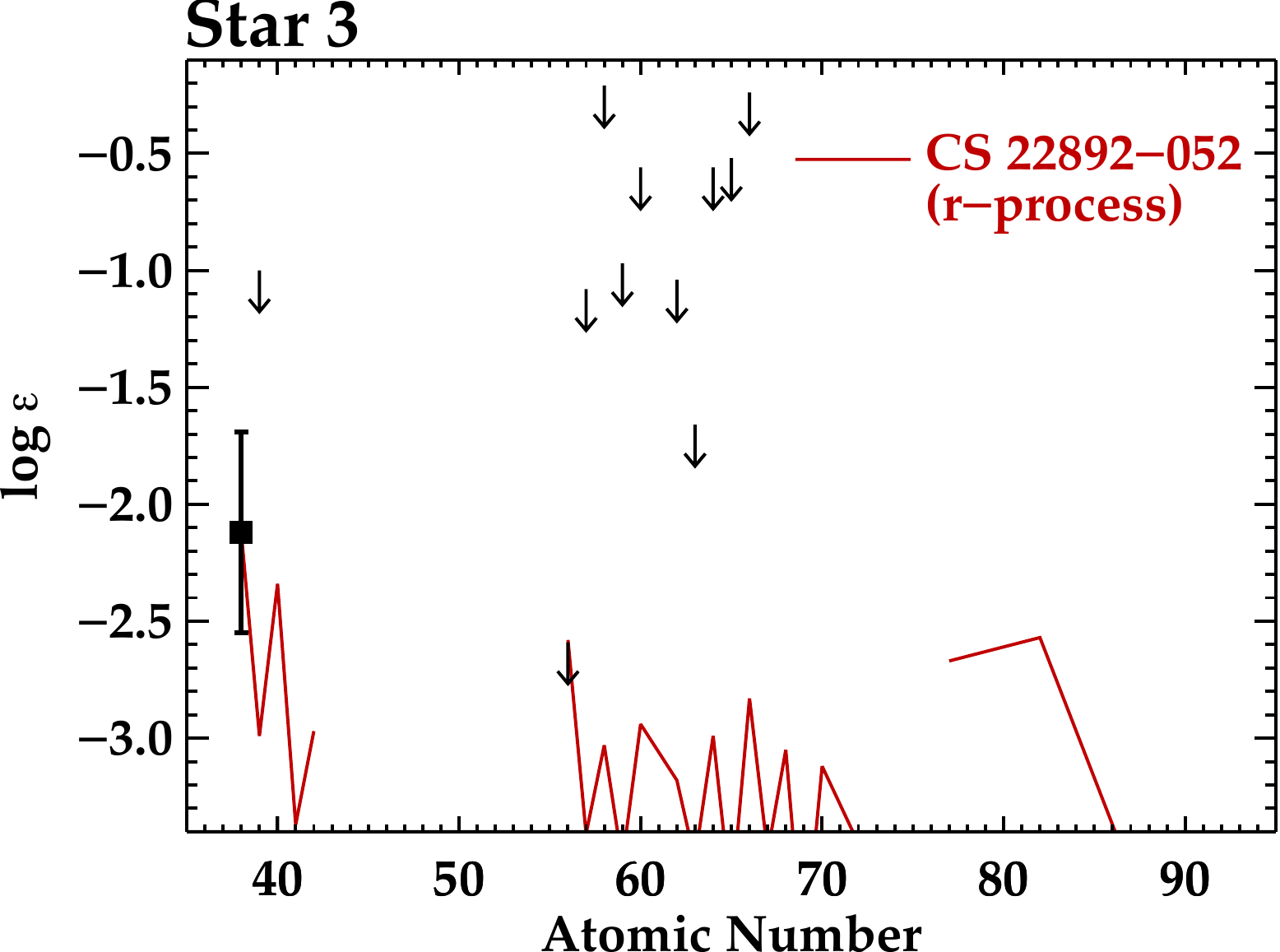}
\hspace*{0.1in}
\includegraphics[angle=0,width=2.60in]{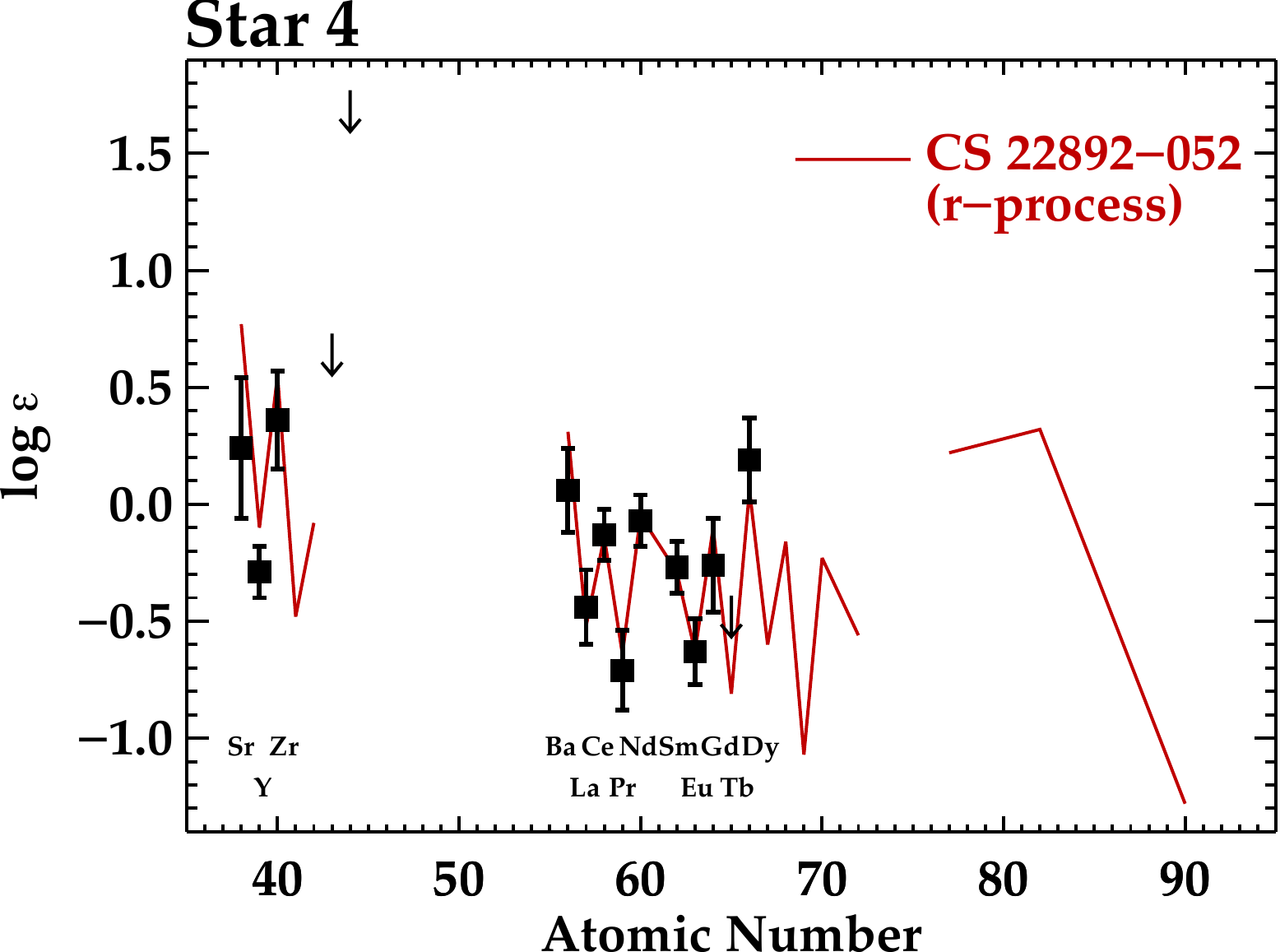} 
\end{center}
\caption{
\label{ncapplot}
The \ncap\ abundance patterns compared with
the \rpro-enhanced standard star \mbox{CS~22892--052}
\citep{sneden03,sneden09,roederer09b}.
The patterns are normalized to Eu ($Z =$~63)
in \stara, \starb, and \stard\
and to Sr ($Z =$~38) in \starc.
Note the different scale on the vertical axis
for \starc\ (lower left panel).
}
\end{figure*}

Figure~\ref{ncapplot}
illustrates the detailed abundance patterns of
each of the four stars observed in \ret.
The abundance pattern found in star \cs\
is shown for comparison.
The only adjustment made to this pattern
is its overall normalization.
This pattern is a superb representation
for the elements detected in \stara,
\starb, and \stard\ in \ret.
The ratio between the lighter \ncap\
elements (Sr, Y, and Zr; 38~$\leq Z \leq$~40)
and the heavier ones (Ba and beyond; $Z \geq$~56)
is also well-matched,
which is not always the case in 
metal-poor stars with \rpro\ material
(e.g., \citealt{mcwilliam98,johnson02b,aoki05,roederer10}).
This comparison also suggests that the
radioactive actinides $^{232}$Th 
and $^{238}$U may be
present and detectable in \ret,
which could offer an independent 
check of the age of the \rpro\ material in this system
(cf., e.g., \citealt{cowan97,cayrel01}), but
no lines of these elements are covered in our spectra.
We conclude that \rpro\ nucleosynthesis
is responsible for the production 
of the heavy elements observed in \ret.

\subsection{Carbon}
\label{carbon}

The [C/Fe] ratios in all four stars observed in \ret\
are super-solar, as shown in Figure~\ref{multiplot}.
However, 
C is depleted at the surface as stars ascend the red giant branch.
We use the corrections given by \citet{placco14}
to estimate the natal [C/Fe] ratios for these stars.
These are listed in Table~\ref{ctab}.
The magnitude of the corrections increases 
with decreasing \logg, as expected.
\starb\ is classified as a C-enhanced metal-poor (CEMP)
star using a common definition of C-enhancement,
[C/Fe]~$> +$0.7 \citep{aoki07a}.
The enhanced C in \starb\ is apparent 
from the strong absorption near 4310
and 4323~\AA\ in the top panel of Figure~\ref{m2fsrangeplot}.

\begin{deluxetable}{cccc}
\tablecaption{Corrected [C/Fe] ratios
\label{ctab}}
\tablewidth{0pt}
\tabletypesize{\scriptsize}
\tablehead{
\colhead{Star} &
\colhead{\logg} &
\colhead{[C/Fe]} &
\colhead{[C/Fe]} \\
\colhead{} &
\colhead{} &
\colhead{(observed)} &
\colhead{(corrected)} 
}
\startdata
Star 1 & 2.09 & $+$0.25 & $+$0.27 \\
Star 2 & 1.22 & $+$0.60 & $+$1.15 \\
Star 3 & 1.63 & $+$0.13 & $+$0.38 \\
Star 4 & 1.50 & $+$0.26 & $+$0.68 
\enddata
\end{deluxetable}

The fraction of CEMP stars
in the field
increases with decreasing metallicity
(e.g., \citealt{frebel06,carollo12,lee13b,placco14}).
CEMP stars with no enhancement of slow \ncap\ process (\spro) material,
like those in \ret,
likely reflect their natal composition
(e.g., \citealt{ryan05,norris13,starkenburg14,hansen15a,hansen15c}).
These CEMP stars
may contain metals from 
one or a few zero-metallicity 
Pop~III stars
(cf., e.g., \citealt{bromm03,umeda05,cooke14,ishigaki14b}),
although they are not the only class of stars
proposed to have formed from the remnants of Pop~III stars
(e.g., \citealt{aoki14}).
A few CEMP stars have been identified in the
UFD galaxies
\citep{frebel10,norris10b,lai11,gilmore13}, 
and
we add one star in \ret\ to this small but growing inventory.

The CEMP star in \ret\ is also substantially 
enhanced in \rpro\ material
(\cempr, according to the nomenclature of \citealt{beers05}).
Only two other stars, \cs\ \citep{sneden94,sneden96} and
\object[BPS CS 22945-017]{CS~22945--017} \citep{roederer14e}
are members of the \cempr\ class.
No compelling explanation exists for the
enhancement of both C and \rpro\ material
in these stars.
Not all \rpro\ stars in \ret\ are C-enhanced.
We infer from this that 
the C and \rpro\ enhancements may not
be co-produced by the same nucleosynthesis
site or mechanism.
By extension, perhaps the C- and \rpro-enhancements in 
other \cempr\ stars
are also not directly related,
as previous studies have concluded.

\subsection{Magnesium through Zinc}
\label{light}

Figure~\ref{multiplot} 
also compares the abundance ratios among
$\alpha$- and Fe-group elements in \ret\ with
those in other UFD galaxies and halo giants.
In all cases, the \ret\ abundance ratios
fall well within the range 
found for other UFD galaxies and halo stars.

\begin{figure}
\begin{center}
\includegraphics[angle=0,width=3.4in]{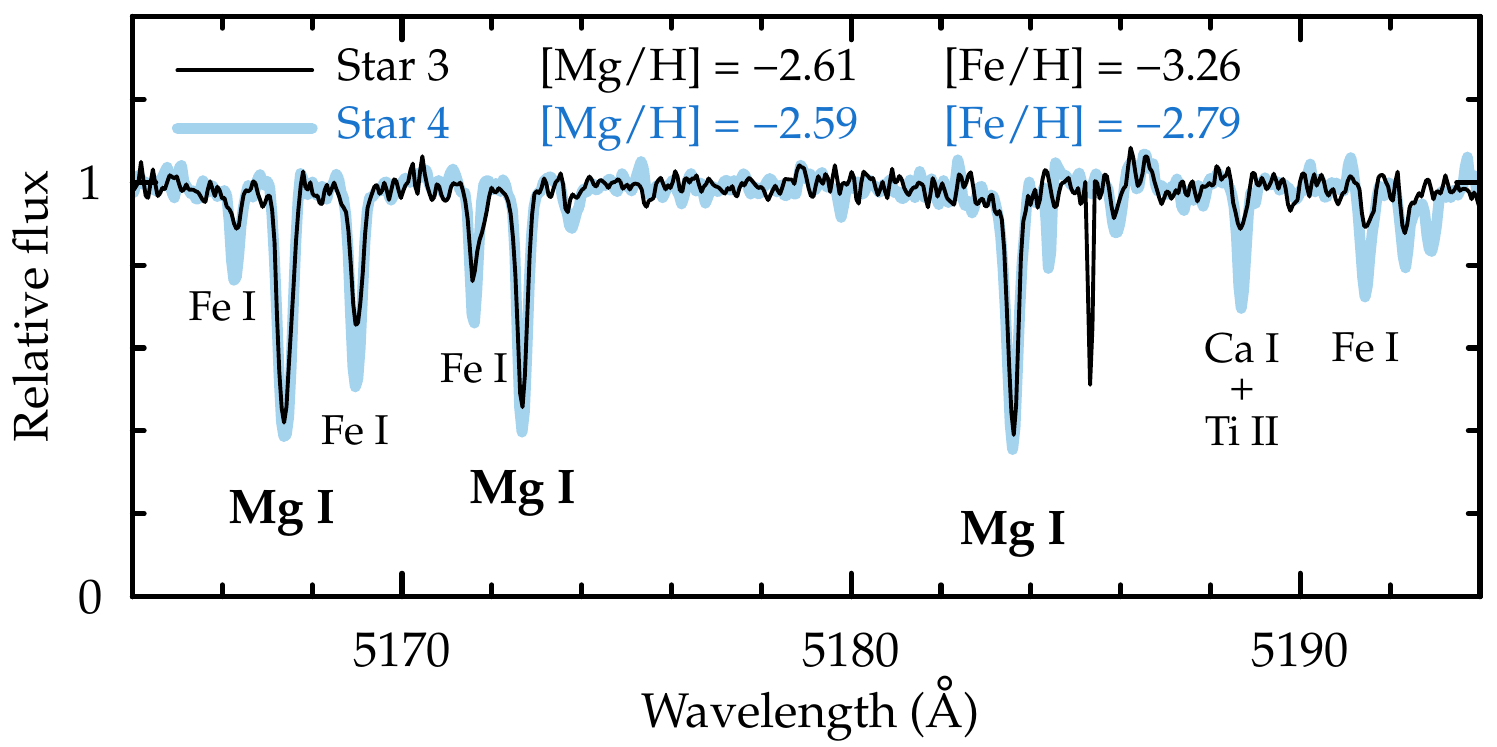}
\end{center}
\caption{
\label{mgplot}
Spectra of Stars~3 and 4 around the Mg~\textsc{i}
lines at 5167.32, 5172.68, and 5183.60~\AA.
Both stars have similar \teff\ and \logg.
The Fe~\textsc{i} lines are significantly weaker
in Star~3, yet
the Mg~\textsc{i} lines have nearly equal strengths.
}
\end{figure}

The [Mg/Fe] ratio is higher in \starc,
the most metal-poor star in \ret\ observed by us, 
by $\approx$~0.45~dex compared to the other three stars.
This difference is about three times larger
than the measurement uncertainties.
Figure~\ref{mgplot} illustrates portions of the
spectra around several Mg~\textsc{i} lines
in \starc\ and \stard, which have
different [Fe/H] but 
similar [Mg/H] ratios.
This comparison supports our assertion that the [Mg/Fe] ratio
is enhanced in \starc.
Enhanced [Mg/Fe] appears to be a genuine characteristic of 
some stars in the UFD galaxies.
Two of these Mg-enhanced stars (in \booonegal\ and \segonegal)
are also C-enhanced
\citep{norris10b,lai11,gilmore13},
but the one in \ret\ and 
several others in \comgal, \hergal, and \umagal\ are not
\citep{koch08,frebel10}.
The enhanced [Mg/Fe] ratios in the C-normal
stars have been interpreted (e.g., \citeauthor{koch08})
as evidence for enrichment dominated by
high-mass core-collapse supernovae.

[Ti~\textsc{i}/Fe] is high in three of the four stars.
The Ti~\textsc{i} abundance has been derived
from 4--12~lines in these three stars.
No line appears to yield systematically high
abundances in \ret, and \citet{roederer14c}
did not identify any of these lines
as yielding systematically-high abundances
in metal-poor halo giants.
This indicates that the high [Ti~\textsc{i}/Fe] ratios
cannot be attributed to a handful of 
blended or poorly-measured Ti~\textsc{i} lines.
We note that 
[Ti~\textsc{ii}/Fe] is not high, however,
suggesting that there is probably not a
genuine excess of Ti in the \ret\ stars.

The [X/Fe] ratios
show minimal or no dispersion within \ret,
except for C, Mg, and the \ncap\ elements
discussed previously.
Furthermore, none of the [Mg/Fe], [Ca/Fe], or [Ti/Fe] ratios
lie significantly below the 
$\alpha$-enhanced plateau 
([$\alpha$/Fe]~$\sim +$0.3) 
in any of the \ret\ stars observed.

\subsection{Outliers in the [X/Fe] Ratios}
\label{outliers}

\begin{figure*}
\begin{center}
\includegraphics[angle=0,width=2.60in]{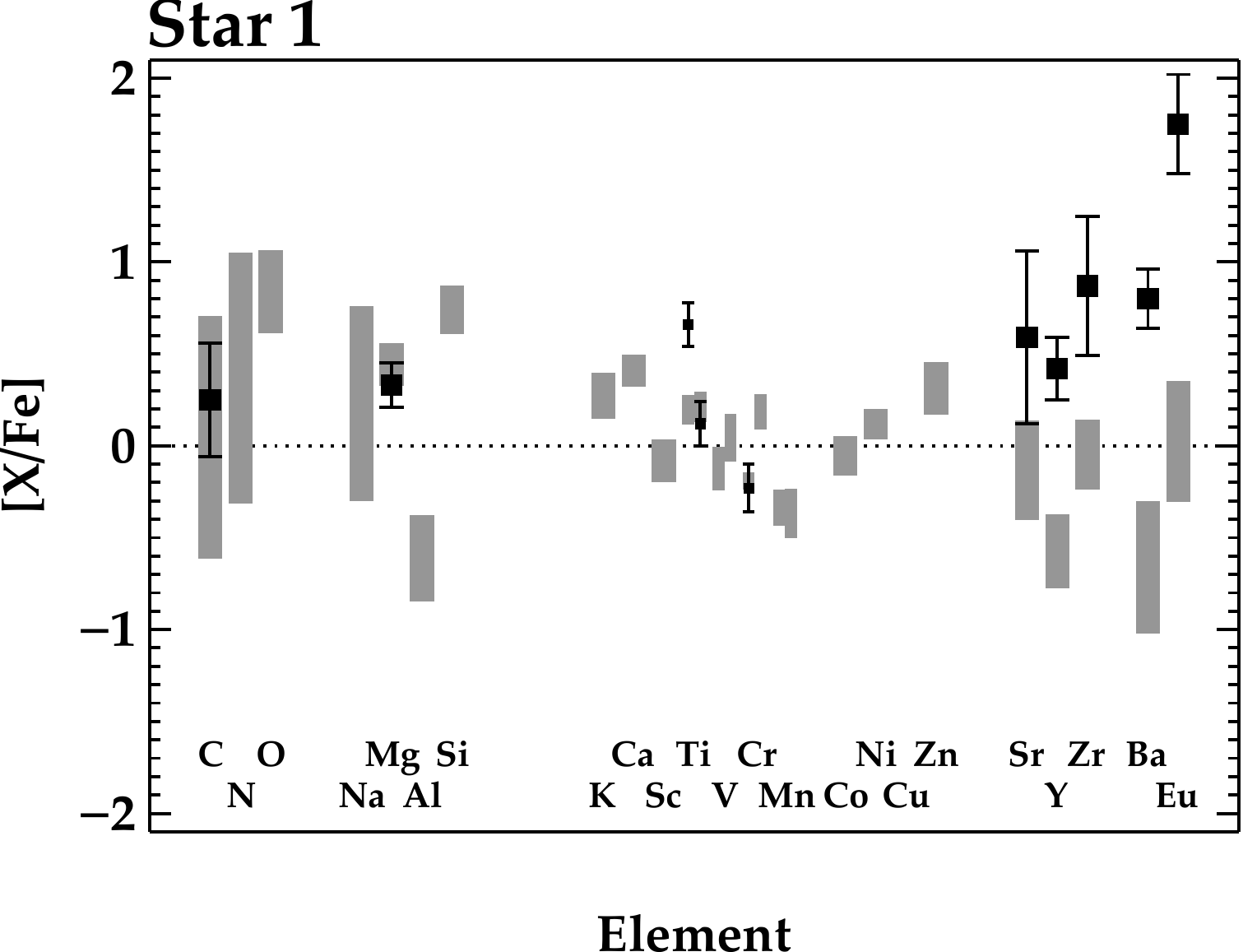}
\hspace*{0.1in}
\includegraphics[angle=0,width=2.60in]{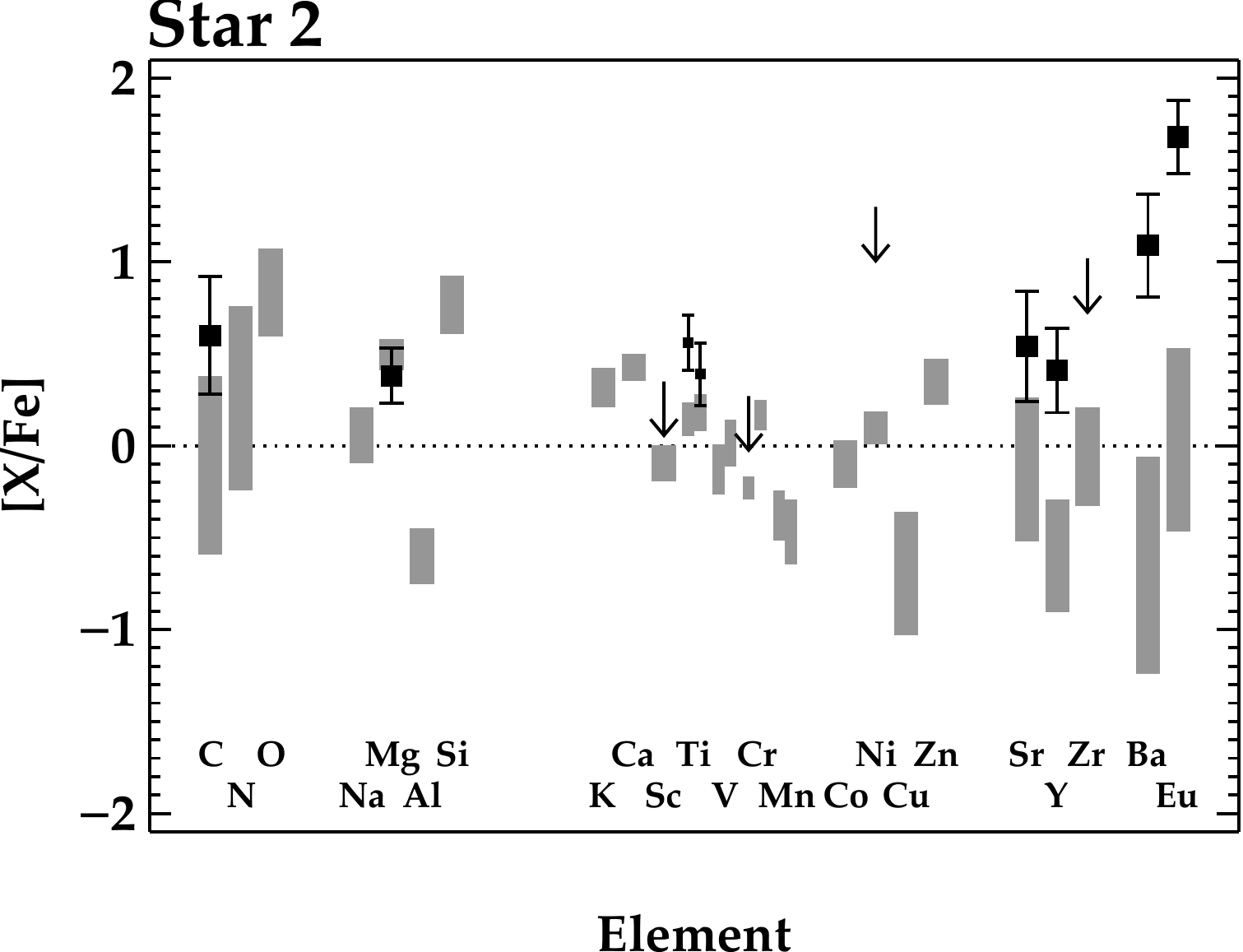} \\
\vspace*{0.1in}
\includegraphics[angle=0,width=2.60in]{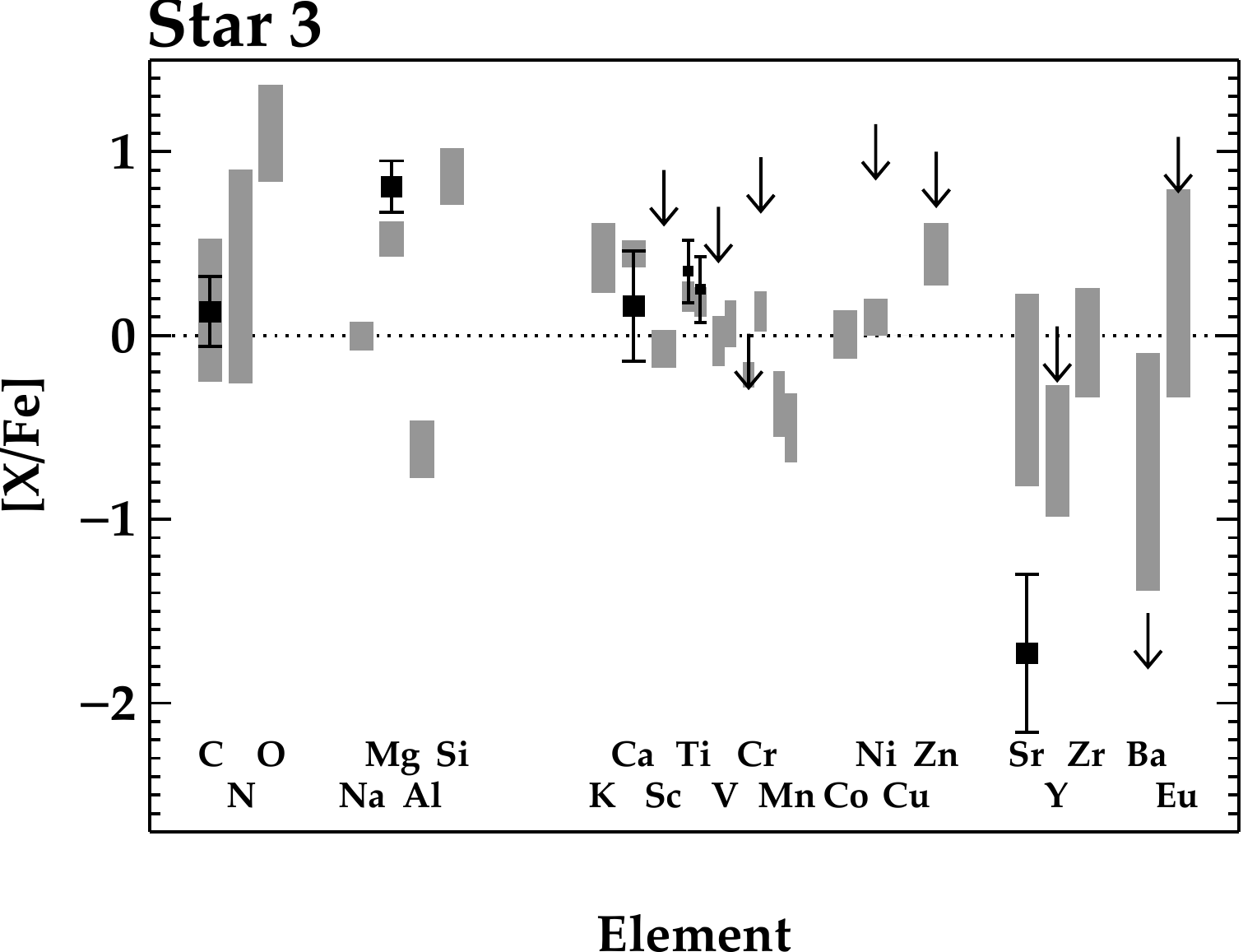}
\hspace*{0.1in}
\includegraphics[angle=0,width=2.60in]{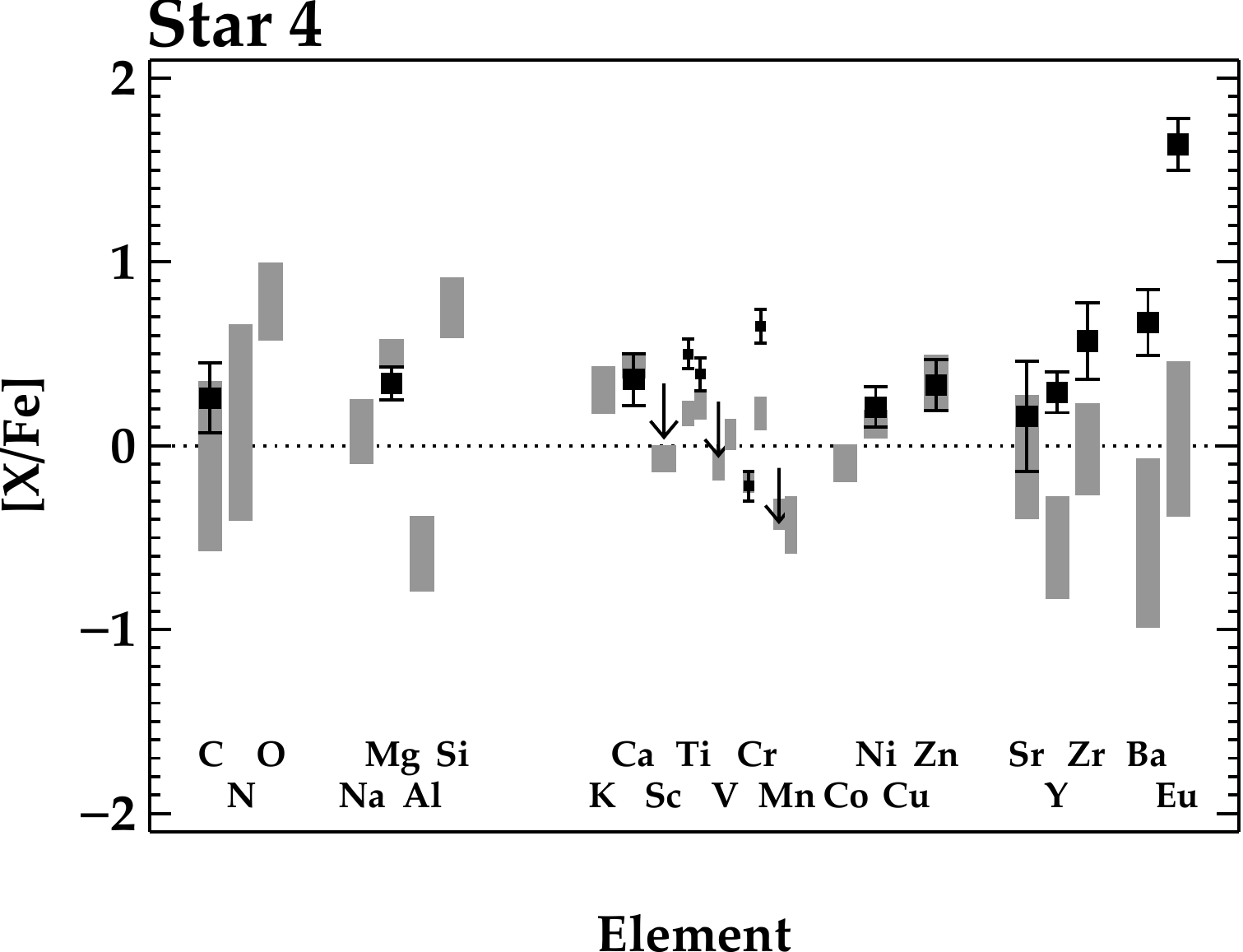} 
\end{center}
\caption{
\label{outlierplot}
Comparison of the abundances in the \mbox{Ret~2} stars
with field red giants with similar metallicities.
The comparison samples, drawn from \citet{roederer14c},
are selected to have \teff\ within 250~K and [Fe/H]
within 0.3~dex of each \mbox{Ret~2} star.
The number of stars in the comparison samples range
from 22 to 34.
This comparison sample, shown by the shaded gray boxes,
represents the mean $\pm$~1$\sigma$ standard deviations.
Smaller symbols are shown for Ti, V, Cr, and Mn
to accommodate ratios from both the neutral and ionized 
states, which may differ.
The dotted line marks the solar ratios.
}
\end{figure*}

In Figure~\ref{outlierplot}, 
we perform a detailed quantitative comparison between
the abundance ratios in 
each of the stars in \ret\
and halo giants with similar \teff\
and [Fe/H] that have been analyzed using similar methods.
This comparison allows us to identify any
element ratios that are outliers
relative to the majority of
halo stars.
The systematic abundance uncertainties largely
cancel out when using this differential approach.
Elements detected in multiple ionization states
are compared and illustrated separately in Figure~\ref{outlierplot}.
We exclude from the comparison sample any stars that 
exhibit high levels of \spro\ enrichment,
since this characteristic frequently signals
that the present-day surface composition of the star
has been polluted by a companion and 
probably does not reflect the natal composition.

The heaviest elements are, of course, significantly overabundant
in \stara, \starb, and \stard\
relative to the comparison sample.
In \starc, the Sr and Ba abundances are not only 
deficient relative to the other stars in \ret, but they
are also deficient relative to
the comparison halo sample.
This deficiency is, however, quite normal
for stars in the UFD galaxies.
\citet{frebel14} suggested that this
may be a result of a single \ncap\ enrichment event within each galaxy.
We can detect only Sr in \starc,
limiting our ability to reliably
identify the nucleosynthetic process 
that produced the heaviest elements in \starc.
Candidates include
the weak component of the $r$~process (e.g., \citealt{truran02,arcones13})
and charged particle reactions (e.g., \citealt{woosley92})
operating in core-collapse supernovae
or 
the weak component of the $s$~process (e.g., \citealt{raiteri91,pignatari08})
operating in massive, rapidly-rotating stars.
The upper limit on Ba in \starc\ (Figure~\ref{ncapplot})
is nearly able to exclude the possibility
that the \rpro\ pattern 
found in the \rpro-enhanced stars 
is also present in \starc\ at an abundance
three orders of magnitude lower.
Higher quality spectra of this star could settle the issue.
Regardless, it seems likely that at least two \ncap\ 
enrichment events occurred in \ret, 
which distinguishes it from other UFD galaxies.

None of the other [X/Fe] ratios in Figure~\ref{outlierplot}
differ by more than 2$\sigma$
between the \ret\ stars and the comparison sample.
We conclude that the $\alpha$ and Fe-group elements
studied in \ret\ were produced in similar proportions
by the progenitors responsible for enriching \ret\ 
and the stars in the halo comparison sample.

\citet{roederer14c} performed this comparison
for stars in the \rtwo\ class.
That study concluded that the abundances of 
elements from Mg to Zn in the \rtwo\ stars
are indistinguishable from those in stars
with normal levels of \rpro\ material
at the limit of the data,
about 3.5\% (0.015~dex).
This result was interpreted to mean that
the nucleosynthetic sites responsible for the 
large \rpro\ enhancements in the \rtwo\ stars
did not produce any $\alpha$- or Fe-group element
abundance signatures distinct from normal
core-collapse supernovae.
Alternatively, the \rpro\ nucleosynthetic site(s)
produced no $\alpha$ or Fe-group elements at all.
We propose that this conclusion
also applies to the stars examined in \ret.

\section{Discussion and Conclusions}
\label{discussion}

The present-day stellar mass of \ret\ 
($\sim$~2.6$\times$10$^{3}$~\msun; \citealt{bechtol15})
is roughly two orders of magnitude less
than the minimum mass necessary to 
fully sample a standard IMF.~
Assuming that \ret\ still retains a 
substantial fraction of its stars,
we would expect to observe the effects of
stochastically sampling the IMF.~
This evidence may be found in the
enhanced [C/Fe] ratio in \starb\ 
and enhanced [Mg/Fe] ratio in \starc.
The normal $\alpha$- and Fe-group abundances
found in most stars offer no compelling evidence of 
substantial variations at the high end of the IMF in \ret.

The UFD enrichment models of \citet{lee13a}
rely on stochastic sampling of the mass function of
supernovae whose 
\ncap\ yields
are more strongly mass-dependent than 
yields of lighter elements like Ti.
These models predict that some
\ncap-rich stars
should be found in the UFDs,
and the \rpro-enhanced stars in \ret\ confirm this prediction.
Neutron-star mergers (e.g., \citealt{goriely11,ramirezruiz15})
or rare \rpro\ events associated with core-collapse supernovae,
like magnetically-induced jets from the proto neutron star 
(e.g., \citealt{fujimoto08,nishimura15}),
can produce a few 10$^{-2}$~\msun\ of \rpro\ material and
a few 10$^{-5}$~\msun\ of Eu.
These are compatible with the observations of \ret,
and the sites associated with supernovae
fulfill the requirement of the \citeauthor{lee13a}\ model.

We can use \ret\ and other UFDs 
to estimate the amount of
material produced by the main component of the \rpro\
per unit of gas turned into stars.
The 10 UFD galaxies whose stars
have been observed at high spectral resolution
(see references in caption to Figure~\ref{multiplot})
have a combined stellar mass of 
$\sim$~1.1$\times$10$^{5}$~\msun\
\citep{mcconnachie12,bechtol15}.
\citet{ji15b} estimated that 
$\sim$~3$\times$10$^{-5}$~\msun\ of Eu
($\sim$~3$\times$10$^{-2}$~\msun\ of \rpro\ material)
are required to match the observations of \ret.
Thus $\sim$~3$\times$10$^{-10}$~\msun\ of Eu
($\sim$~3$\times$10$^{-7}$~\msun\ of \rpro\ material)
is produced per \msun\ of gas turned into stars in the
10 UFD galaxies.

This rate may offer 
a link between the UFDs and higher-mass dSph galaxies.
\citet{tsujimoto15} identified a plateau at [Eu/H]~$= -$1.3
in stars with [Fe/H]~$\gtrsim -$2
in \dragal\ and several other dSph galaxies.
\dragal\ has a total stellar mass of $\sim$~2.5$\times$10$^{5}$~\msun\
in its stars with [Fe/H]~$> -$2 (\citeauthor{tsujimoto15}),
which translates to $\sim$~5$\times$10$^{-5}$~\msun\
of Eu.
Using the rate of Eu production we estimate based on the UFD galaxies,
the same stellar mass in \dragal\ would
be expected to produce $\sim$~7$\times$10$^{-5}$~\msun\ of Eu.
The similarity between these two values could
reflect a natural connection between the
\rpro\ material
observed in surviving UFDs and dSph galaxies.
Material produced by the main component of the \rpro\ 
has also been found in \cargal\ and \umigal,
two other dSph galaxies
with stellar masses similar to \dragal\
\citep{shetrone01,shetrone03,sadakane04,cohen10,venn12}.
The extreme \rpro\ enhancement observed in \ret\ 
may be the result of 
a low-probability nucleosynthesis event 
that becomes inevitable (and less conspicuous)
in more luminous dSph galaxies with stellar masses
$\gtrsim$~10$^{5}$~\msun.
\citet{tafelmeyer10} and \citet{jablonka15} 
called for examination of the
neutron-capture abundance patterns in metal-poor stars in dSph galaxies
with a range of masses 
to better understand the galactic metal retention and 
(Pop~II or Pop~III) sites that produce heavy elements.
We propose that a focused effort to quantify
the fraction of metal-poor stars
with material produced by the main component of the \rpro\
as a function of galaxy mass may also prove enlightening.

\citet{frebel12} identified a set of chemical
characteristics that could distinguish a
bona fide ``first galaxy,''
a system where only one 
long-lived stellar generation 
formed from the yields of the first Pop~III stars.
\ret\ fulfills some of the observational criteria
identified by \citeauthor{frebel12},
since there is no evidence of enrichment by 
Type~Ia supernovae or asymptotic giant branch stars.
However, \ret\ appears not to be a good candidate
``first galaxy'' based on the
large internal spread in \rpro\ material.
The larger sample of \citet{ji15b} hints
that chemical evolution may have occurred
within \ret,
since the most metal-poor stars lack the
\rpro\ enhancements observed in the
more metal-rich ones.

The escape velocities of the UFD systems like \ret\
are $\sim$~25--50~\kmsec, 
assuming a NFW halo profile \citep{navarro96}
with virial mass $\sim$~10$^{8}$--10$^{9}$~\msun\ 
and a concentration, $c$, of 10.7--12.5 \citep{prada12}.
Most ($\gg$~99\%) of the metals produced within 
\ret\ were probably lost from the system,
as has been inferred for other dwarf galaxies
(e.g., \citealt{kirby11,frebel14}).
Alternatively, simulations by \citet{smith15} indicate that
some low-mass minihalos
could have been externally enriched by
metals from neighboring minihalos 
prior to formation of their own Pop~III star.
The star-to-star abundance spreads observed in some element ratios
(e.g., [Fe/H], [C/Fe], [Mg/Fe], [Eu/Fe])
indicate that the ISM of \ret\ was not fully mixed
at the time the low-mass stars formed.
This indicates that ejecta from
at least two supernovae---perhaps even from
multiple neighboring minihalos---formed stars 
in the minihalo that evolved into \ret.
Simulations tailored specifically to reproduce the
conditions and chemistry in \ret\ 
(cf., e.g., \citealt{ritter15,webster15})
may prove illuminating on these points.

\acknowledgments

We express our deep appreciation to the referee for offering a
positive, constructive, and timely review.
I.U.R.\ acknowledges partial support 
from grant PHY~14-30152 (Physics Frontier Center/JINA-CEE)
awarded by the U.S.\ National Science Foundation (NSF).~
M.M.\ and J.I.B.\ gratefully acknowledge support
from NSF grant AST~09-23160
to develop M2FS.~
S.R.L.\ acknowledges the Michigan Society of Fellows for financial support.
D.L.N.\ was supported by a McLaughlin Fellowship at the University of Michigan.
E.O.\ is partially supported by NSF grant AST~13-13006.
M.V.\ is supported in part by 
HST-AR-13890.001, NSF grant AST-09-08346, 
and NASA-ATP award NNX15AK79G.~
M.G.W.\ is supported by NFS grants AST~13-13045 and AST~14-12999.
This research has made use of NASA's
Astrophysics Data System Bibliographic Services;
the arXiv pre-print server operated by Cornell University;
the SIMBAD 
database hosted by the
Strasbourg Astronomical Data Center;
the ASD hosted by NIST;
IRAF software packages
distributed by the National Optical Astronomy Observatories,
which are operated by the Association of Universities for Research
in Astronomy, Inc., under cooperative agreement with the National
Science Foundation;
and the R software package \citep{r}.

{\it Facilities:} 
\facility{Magellan(M2FS)}


\begin{thebibliography}{}


\bibitem[Aldenius et al.(2007)]{aldenius07} Aldenius, M., Tanner, J.~D., 
Johansson, S., Lundberg, H., Ryan, S.~G.\ 2007, \aap, 461, 767 

\bibitem[Aoki et al.(2005)]{aoki05} Aoki, W., Honda, S., 
Beers, T.~C., et al.\ 2005, \apj, 632, 611 

\bibitem[Aoki et al.(2007a)]{aoki07a} Aoki, W., Beers, T.~C., 
Christlieb, N., et al.\ 2007a, \apj, 655, 492 

\bibitem[Aoki et al.(2007b)]{aoki07b} Aoki, W., Honda, S., 
Sadakane, K., Arimoto, N.\ 2007b, \pasj, 59, L15 

\bibitem[Aoki et al.(2014)]{aoki14} Aoki, W., Tominaga, N., 
Beers, T.~C., Honda, S., Lee, Y.~S.\ 2014, Science, 345, 912 

\bibitem[Arcones \& Thielemann(2013)]{arcones13} Arcones, A., 
Thielemann, F.-K.\ 2013, J.\ Physics G Nuclear Physics, 40, 013201 

\bibitem[Argast et al.(2004)]{argast04} Argast, D., Samland, M., 
Thielemann, F.-K., Qian, Y.-Z.\ 2004, \aap, 416, 997 

\bibitem[Asplund et al.(2009)]{asplund09} Asplund, M., Grevesse, N., 
Sauval, A.~J., Scott, P.\ 2009, \araa, 47, 481 

\bibitem[Bailey et al.(2012)]{bailey12} Bailey, J.~I., Mateo, 
M.~L., Bagish, A.~P., Crane, J.~D., 
Slater, C.~T.\ 2012, \procspie, 8446, 84465G 

\bibitem[Barklem et al.(2005)]{barklem05} Barklem, P.~S., 
Christlieb, N., Beers, T.~C., et al.\ 2005, \aap, 439, 129 

\bibitem[Bechtol et al.(2015)]{bechtol15} Bechtol, K., 
Drlica-Wagner, A., Balbinot, E., et al.\ 2015, \apj, 807, 50 

\bibitem[Bedell et al.(2014)]{bedell14} Bedell, M., 
Mel{\'e}ndez, J., Bean, J.~L., et al.\ 2014, \apj, 795, 23 

\bibitem[Beers \& Christlieb(2005)]{beers05} Beers, T.~C., 
Christlieb, N.\ 2005, \araa, 43, 531 

\bibitem[Bi{\'e}mont et al.(2011)]{biemont11} Bi{\'e}mont, 
 {\'E}., Blagoev, K., Engstr{\"o}m, L., et al.\ 2011, \mnras, 414, 3350 

\bibitem[Booth et al.(1984)]{booth84} Booth, A.~J., Blackwell, 
D.~E., Petford, A.~D., Shallis, M.~J.\ 1984, \mnras, 208, 147 

\bibitem[Bromm \& Loeb(2003)]{bromm03} Bromm, V., Loeb, A.\ 
2003, \nat, 425, 812 

\bibitem[Bromm \& Larson(2004)]{bromm04} Bromm, V., 
Larson, R.~B.\ 2004, \araa, 42, 79 

\bibitem[Brown et al.(2012)]{brown12} Brown, T.~M., Tumlinson, 
J., Geha, M., et al.\ 2012, \apjl, 753, L21 

\bibitem[Carollo et al.(2012)]{carollo12} Carollo, D., Beers, 
T.~C., Bovy, J., et al.\ 2012, \apj, 744, 195 

\bibitem[Castelli \& Kurucz(2003)]{castelli03} Castelli, F., Kurucz, R.~L.\
Proc.\ IAU Symp.\ No 210, Modelling of Stellar Atmospheres,
N.\ Piskunov et al., eds.\ 2003, A20

\bibitem[Cayrel et al.(2001)]{cayrel01} Cayrel, R., Hill, V., 
Beers, T.~C., et al.\ 2001, \nat, 409, 691 

\bibitem[Cescutti et al.(2015)]{cescutti15} Cescutti, G., Romano, D., 
Matteucci, F., Chiappini, C., Hirschi, R.\ 2015, \aap, 577, A139 

\bibitem[Christlieb et al.(2004)]{christlieb04} Christlieb, N., 
Beers, T.~C., Barklem, P.~S., et al.\ 2004, \aap, 428, 1027 

\bibitem[Cohen \& Huang(2010)]{cohen10} Cohen, J.~G., 
Huang, W.\ 2010, \apj, 719, 931 

\bibitem[Cooke \& Madau(2014)]{cooke14} Cooke, R.~J., 
Madau, P.\ 2014, \apj, 791, 116 

\bibitem[Cowan et al.(1997)]{cowan97} Cowan, J.~J., McWilliam, 
A., Sneden, C., Burris, D.~L.\ 1997, \apj, 480, 246 

\bibitem[Creevey et al.(2012)]{creevey12} Creevey, O.~L., 
Th{\'e}venin, F., Boyajian, T.~S., et al.\ 2012, \aap, 545, A17 

\bibitem[Demarque et al.(2004)]{demarque04} Demarque, P., Woo, 
J.-H., Kim, Y.-C., Yi, S.~K.\ 2004, \apjs, 155, 667 

\bibitem[Den Hartog et al.(2003)]{denhartog03} Den Hartog, E.~A., 
Lawler, J.~E., Sneden, C., Cowan, J.~J.\ 2003, \apjs, 148, 543 

\bibitem[Den Hartog et al.(2006)]{denhartog06} Den Hartog, E.~A., 
Lawler, J.~E., Sneden, C., Cowan, J.~J.\ 2006, \apjs, 167, 292 

\bibitem[Den Hartog et al.(2011)]{denhartog11} Den Hartog, E.~A., 
Lawler, J.~E., Sobeck, J.~S., Sneden, C., 
Cowan, J.~J.\ 2011, \apjs, 194, 35 

\bibitem[Diehl et al.(2014)]{diehl14} Diehl, H.~T., Abbott, 
T.~M.~C., Annis, J., et al.\ 2014, \procspie, 9149, 91490V 

\bibitem[Drlica-Wagner et al.(2015)]{drlicawagner15} Drlica-Wagner, 
A., Bechtol, K., Rykoff, E.~S., et al.\ 2015, \apj, 813, 109 

\bibitem[Feltzing et al.(2009)]{feltzing09} Feltzing, S., 
Eriksson, K., Kleyna, J., Wilkinson, M.~I.\ 2009, \aap, 508, L1 

\bibitem[Fran\c{c}ois et al.(2015)]{francois15} Fran\c{c}ois, 
P., Monaco, L., Bonifacio, P., et al.\ 2015, \aap, submitted (arXiv:1510.05401)

\bibitem[Frebel et al.(2006)]{frebel06} Frebel, A., Christlieb, 
N., Norris, J.~E., et al.\ 2006, \apj, 652, 1585 

\bibitem[Frebel et al.(2007)]{frebel07} Frebel, A., Christlieb, 
N., Norris, J.~E., et al.\ 2007, \apjl, 660, L117 

\bibitem[Frebel et al.(2010)]{frebel10} Frebel, A., Simon, 
J.~D., Geha, M., Willman, B.\ 2010, \apj, 708, 560 

\bibitem[Frebel \& Bromm(2012)]{frebel12} Frebel, A., Bromm, V.\ 
2012, \apj, 759, 115 

\bibitem[Frebel et al.(2013)]{frebel13} Frebel, A., Casey, 
A.~R., Jacobson, H.~R., Yu, Q.\ 2013, \apj, 769, 57 

\bibitem[Frebel et al.(2014)]{frebel14} Frebel, A., Simon, 
J.~D., Kirby, E.~N.\ 2014, \apj, 786, 74 

\bibitem[Fujimoto et al.(2008)]{fujimoto08} Fujimoto, S.-i., 
Nishimura, N., \& Hashimoto, M.-a.\ 2008, \apj, 680, 1350 

\bibitem[Geringer-Sameth et al.(2015)]{geringersameth15} 
Geringer-Sameth, A., Walker, M.~G., Koushiappas, S.~M., et al.\ 2015, 
\prl, 115, 081101 

\bibitem[Gilmore et al.(2013)]{gilmore13} Gilmore, G., Norris, 
J.~E., Monaco, L., et al.\ 2013, \apj, 763, 61 

\bibitem[Goriely et al.(2011)]{goriely11} Goriely, S., Bauswein, 
A., Janka, H.-T.\ 2011, \apjl, 738, L32 

\bibitem[Hansen et al.(2011)]{hansen11} Hansen, T., Andersen, 
J., Nordstr{\"o}m, B., Buchhave, L.~A., 
Beers, T.~C.\ 2011, \apjl, 743, L1 

\bibitem[Hansen et al.(2015a)]{hansen15a} Hansen, T., Hansen, 
C.~J., Christlieb, N., et al.\ 2015a, \apj, 807, 173 

\bibitem[Hansen et al.(2015b)]{hansen15b} Hansen, T.~T., Andersen, J., 
Nordstr{\"o}m, B., et al.\ 2015b, \aap, 583, A49 

\bibitem[Hansen et al.(2015c)]{hansen15c} Hansen, T.~T., Andersen, 
J., Nordstr{\"o}m, B., et al.\ 2015c, \aap, in press (arXiv:1511.08197)

\bibitem[Hayek et al.(2009)]{hayek09} Hayek, W., Wiesendahl, U., 
Christlieb, N., et al.\ 2009, \aap, 504, 511 

\bibitem[Hill et al.(2002)]{hill02} Hill, V., Plez, B., 
Cayrel, R., et al.\ 2002, \aap, 387, 560 

\bibitem[Hotokezaka et al.(2015)]{hotokezaka15} Hotokezaka, K., 
Piran, T., Paul, M.\ 2015, Nature Physics, 11, 1042

\bibitem[Ishigaki et al.(2014a)]{ishigaki14a} Ishigaki, M.~N., 
Aoki, W., Arimoto, N., Okamoto, S.\ 2014a, \aap, 562, A146 

\bibitem[Ishigaki et al.(2014b)]{ishigaki14b} Ishigaki, M.~N., 
Tominaga, N., Kobayashi, C., Nomoto, K.\ 2014b, \apjl, 792, L32 

\bibitem[Ishimaru \& Wanajo(1999)]{ishimaru99} Ishimaru, Y., 
Wanajo, S.\ 1999, \apjl, 511, L33 

\bibitem[Ishimaru et al.(2015)]{ishimaru15} Ishimaru, Y., Wanajo, 
S., Prantzos, N.\ 2015, \apjl, 804, L35 

\bibitem[Ivans et al.(2006)]{ivans06} Ivans, I.~I., Simmerer, J., 
Sneden, C., et al.\ 2006, \apj, 645, 613 

\bibitem[Ivarsson et al.(2001)]{ivarsson01} Ivarsson, S., 
Litz{\'e}n, U., Wahlgren, G.~M.\ 2001, \physscr, 64, 455 

\bibitem[Jablonka et al.(2015)]{jablonka15} Jablonka, P., North, P., 
Mashonkina, L., et al.\ 2015, \aap, 583, A67 

\bibitem[Ji et al.(2015a)]{ji15a} Ji, A.~P., Frebel, A., 
Simon, J.~D., Geha, M.\ 2015a, ApJ, submitted (arXiv:1510.07632)

\bibitem[Ji et al.(2015b)]{ji15b} Ji, A.~P., Frebel, A., 
Chiti, A., Simon, J.~D.\ 2015b, Nature, submitted (arXiv:1512.01558)

\bibitem[Johnson(2002)]{johnson02} Johnson, J.~A.\ 2002, \apjs, 139, 219 

\bibitem[Johnson \& Bolte(2002)]{johnson02b} Johnson, J.~A., 
Bolte, M.\ 2002, \apj, 579, 616 

\bibitem[Johnson et al.(2013)]{johnson13} Johnson, C.~I., 
McWilliam, A., Rich, R.~M.\ 2013, \apjl, 775, L27 

\bibitem[Kasen et al.(2015)]{kasen15} Kasen, D., Fern{\'a}ndez, 
R., Metzger, B.~D.\ 2015, \mnras, 450, 1777 

\bibitem[Kirby et al.(2011)]{kirby11} Kirby, E.~N., Martin, 
C.~L., Finlator, K.\ 2011, \apjl, 742, L25 

\bibitem[Koch \& Rich(2014)]{koch14} Koch, A., Rich, R.~M.\ 2014, \apj, 794, 89 

\bibitem[Koch et al.(2008)]{koch08} Koch, A., McWilliam, A., 
Grebel, E.~K., Zucker, D.~B., Belokurov, V.\ 2008, \apjl, 688, L13 

\bibitem[Koch et al.(2013)]{koch13} Koch, A., Feltzing, S., Ad{\'e}n, D., 
Matteucci, F.\ 2013, \aap, 554, A5 

\bibitem[Komiya et al.(2014)]{komiya14} Komiya, Y., Yamada, S., 
Suda, T., Fujimoto, M.~Y.\ 2014, \apj, 783, 132 

\bibitem[Koposov et al.(2015a)]{koposov15a} Koposov, S.~E., 
Belokurov, V., Torrealba, G., Evans, N.~W.\ 2015a, \apj, 805, 130 

\bibitem[Koposov et al.(2015b)]{koposov15b} Koposov, S.~E., Casey, 
A.~R., Belokurov, V., et al.\ 2015b, \apj, 811, 62 

\bibitem[Kramida et al.(2015)]{kramida15} Kramida, A., Ralchenko, Y.,
Reader, J., and the NIST ASD Team.\ 2015, NIST Atomic Spectra Database
 (v.\ 5.3), online, URL:\ http://physics.nist.gov/asd

\bibitem[Kratz et al.(2007)]{kratz07} Kratz, K.-L., Farouqi, 
K., Pfeiffer, B., et al.\ 2007, \apj, 662, 39 

\bibitem[Kurucz \& Bell(1995)]{kurucz95} Kurucz, R.~L., Bell, B.\ 
1995, Kurucz CD-ROM, Cambridge, MA: Smithsonian Astrophysical Observatory

\bibitem[Lai et al.(2011)]{lai11} Lai, D.~K., Lee, Y.~S., 
Bolte, M., et al.\ 2011, \apj, 738, 51 

\bibitem[Lawler \& Dakin(1989)]{lawler89} Lawler, J.~E., Dakin, J.~T.\ 
1989, Journal of the Optical Society of America B Optical Physics, 6, 1457 

\bibitem[Lawler et al.(2001a)]{lawler01a} Lawler, J.~E., 
Bonvallet, G., Sneden, C.\ 2001a, \apj, 556, 452 

\bibitem[Lawler et al.(2001b)]{lawler01b} Lawler, J.~E., 
Wickliffe, M.~E., den Hartog, E.~A., Sneden, C.\ 2001b, \apj, 563, 1075 

\bibitem[Lawler et al.(2001c)]{lawler01c} Lawler, J.~E., 
Wickliffe, M.~E., Cowley, C.~R., Sneden, C.\ 2001c, \apjs, 137, 341 

\bibitem[Lawler et al.(2001d)]{lawler01d} Lawler, J.~E., Wyart, 
J.-F., Blaise, J.\ 2001d, \apjs, 137, 351 

\bibitem[Lawler et al.(2006)]{lawler06} Lawler, J.~E., Den 
Hartog, E.~A., Sneden, C., Cowan, J.~J.\ 2006, \apjs, 162, 227 

\bibitem[Lawler et al.(2009)]{lawler09} Lawler, J.~E., Sneden, 
C., Cowan, J.~J., Ivans, I.~I., Den Hartog, E.~A.\ 2009, \apjs, 182, 51 

\bibitem[Lawler et al.(2013)]{lawler13} Lawler, J.~E., Guzman, 
A., Wood, M.~P., Sneden, C., Cowan, J.~J.\ 2013, \apjs, 205, 11 

\bibitem[Lawler et al.(2014)]{lawler14} Lawler, J.~E., Wood, 
M.~P., Den Hartog, E.~A., et al.\ 2014, \apjs, 215, 20 

\bibitem[Lee et al.(2013)]{lee13a} Lee, D.~M., Johnston, 
K.~V., Tumlinson, J., Sen, B., Simon, J.~D.\ 2013a, \apj, 774, 103 

\bibitem[Lee et al.(2013b)]{lee13b} Lee, Y.~S., Beers, T.~C., 
Masseron, T., et al.\ 2013b, \aj, 146, 132 

\bibitem[Li et al.(2007)]{li07} Li, R., Chatelain, R., Holt, R.~A., 
et al.\ 2007, \physscr, 76, 577 

\bibitem[Ljung et al.(2006)]{ljung06} Ljung, G., Nilsson, H., Asplund, M., 
Johansson, S.\ 2006, \aap, 456, 1181 

\bibitem[Mateo et al.(2012)]{mateo12} Mateo, M., Bailey, J.~I., 
Crane, J., et al.\ 2012, \procspie, 8446, 84464Y 

\bibitem[Mathews et al.(1992)]{mathews92} Mathews, G.~J., Bazan, 
G., Cowan, J.~J.\ 1992, \apj, 391, 719 

\bibitem[Matteucci et al.(2014)]{matteucci14} Matteucci, F., 
Romano, D., Arcones, A., Korobkin, O., Rosswog, S.\ 2014, \mnras, 438, 2177 

\bibitem[McConnachie(2012)]{mcconnachie12} McConnachie, A.~W.\ 
2012, \aj, 144, 4 

\bibitem[McWilliam(1998)]{mcwilliam98} McWilliam, A.\ 1998, \aj, 115, 1640 

\bibitem[McWilliam et al.(1995)]{mcwilliam95} McWilliam, A., 
Preston, G.~W., Sneden, C., Searle, L.\ 1995, \aj, 109, 2757 

\bibitem[Metzger et al.(2010)]{metzger10} Metzger, B.~D., 
Mart{\'{\i}}nez-Pinedo, G., Darbha, S., et al.\ 2010, \mnras, 406, 2650 

\bibitem[Navarro et al.(1996)]{navarro96} Navarro, J.~F., Frenk, 
C.~S., White, S.~D.~M.\ 1996, \apj, 462, 563 

\bibitem[Nilsson et al.(2006)]{nilsson06} Nilsson, H., Ljung, G., 
Lundberg, H., Nielsen, K.~E.\ 2006, \aap, 445, 1165 

\bibitem[Nishimura et al.(2015)]{nishimura15} Nishimura, N., 
Takiwaki, T., Thielemann, F.-K.\ 2015, \apj, 810, 109 

\bibitem[Norris et al.(2010a)]{norris10a} Norris, J.~E., Yong, D., 
Gilmore, G., Wyse, R.~F.~G.\ 2010a, \apj, 711, 350  

\bibitem[Norris et al.(2010b)]{norris10b} Norris, J.~E., Gilmore, 
G., Wyse, R.~F.~G., Yong, D., Frebel, A.\ 2010b, \apjl, 722, L104  

\bibitem[Norris et al.(2013)]{norris13} Norris, J.~E., Yong, D., 
Bessell, M.~S., et al.\ 2013, \apj, 762, 28 

\bibitem[Palmeri et al.(2005)]{palmeri05} Palmeri, P., Fischer, 
C.~F., Wyart, J.-F., Godefroid, M.~R.\ 2005, \mnras, 363, 452 

\bibitem[Pignatari et al.(2008)]{pignatari08} Pignatari, M., 
Gallino, R., Meynet, G., et al.\ 2008, \apjl, 687, L95 

\bibitem[Placco et al.(2014)]{placco14} Placco, V.~M., Frebel, 
A., Beers, T.~C., Stancliffe, R.~J.\ 2014, \apj, 797, 21 

\bibitem[Prada et al.(2012)]{prada12} Prada, F., Klypin, A.~A., 
Cuesta, A.~J., Betancort-Rijo, J.~E., 
Primack, J.\ 2012, \mnras, 423, 3018 

\bibitem[Qian \& Wasserburg(2007)]{qian07} Qian, Y.-Z., 
Wasserburg, G.~J.\ 2007, \physrep, 442, 237 

\bibitem[R Core Team(2014)]{r} R Core Team, 2014.
``R:\ A language and environment for statistical
  computing,'' R Foundation for Statistical Computing, Vienna, Austria.
  URL http://www.R-project.org/.

\bibitem[Raiteri et al.(1991)]{raiteri91} Raiteri, C.~M., Busso, 
M., Picchio, G., Gallino, R., Pulone, L.\ 1991, \apj, 367, 228 

\bibitem[Ramirez-Ruiz et al.(2015)]{ramirezruiz15} Ramirez-Ruiz, E., 
Trenti, M., MacLeod, M., et al.\ 2015, \apjl, 802, L22 

\bibitem[Ritter et al.(2015)]{ritter15} Ritter, J.~S., Sluder, 
A., Safranek-Shrader, C., Milosavljevi{\'c}, M., 
Bromm, V.\ 2015, \mnras, 451, 1190 

\bibitem[Roederer(2009)]{roederer09a} Roederer, I.~U.\ 2009, \aj, 137, 272 

\bibitem[Roederer(2013)]{roederer13} Roederer, I.~U.\ 2013, \aj, 145, 26 

\bibitem[Roederer \& Lawler(2012)]{roederer12a} Roederer, I.~U., 
Lawler, J.~E.\ 2012, \apj, 750, 76 

\bibitem[Roederer \& Kirby(2014)]{roederer14b} Roederer, I.~U., 
Kirby, E.~N.\ 2014, \mnras, 440, 2665 

\bibitem[Roederer et al.(2008a)]{roederer08} Roederer, I.~U., 
Lawler, J.~E., Sneden, C., et al.\ 2008a, \apj, 675, 723 

\bibitem[Roederer et al.(2009)]{roederer09b} Roederer, I.~U., 
Kratz, K.-L., Frebel, A., et al.\ 2009, \apj, 698, 1963 

\bibitem[Roederer et al.(2010)]{roederer10} Roederer, I.~U., 
Cowan, J.~J., Karakas, A.~I., et al.\ 2010, \apj, 724, 975 

\bibitem[Roederer et al.(2014a)]{roederer14a} Roederer, I.~U., 
Preston, G.~W., Thompson, I.~B., Shectman, S.~A., 
Sneden, C.\ 2014a, \apj, 784, 158 

\bibitem[Roederer et al.(2014b)]{roederer14c} Roederer, I.~U., 
Preston, G.~W., Thompson, I.~B., et al.\ 2014c, \aj, 147, 136 

\bibitem[Roederer et al.(2014c)]{roederer14e} Roederer, I.~U., 
Cowan, J.~J., Preston, G.~W., et al.\ 2014c, \mnras, 445, 2946 

\bibitem[Roederer et al.(2016)]{roederer16} Roederer, I.~U., 
Mateo, M., Bailey, J.~I., et al.\ 2016, \mnras, 455, 2417 

\bibitem[Ruffoni et al.(2014)]{ruffoni14} Ruffoni, M.~P., Den 
Hartog, E.~A., Lawler, J.~E., et al.\ 2014, \mnras, 441, 3127 

\bibitem[Ryan et al.(2005)]{ryan05} Ryan, S.~G., Aoki, W., 
Norris, J.~E., Beers, T.~C.\ 2005, \apj, 635, 349 

\bibitem[Sadakane et al.(2004)]{sadakane04} Sadakane, K., Arimoto, 
N., Ikuta, C., et al.\ 2004, \pasj, 56, 1041 

\bibitem[Shetrone et al.(2001)]{shetrone01} Shetrone, M.~D., 
C{\^o}t{\'e}, P., Sargent, W.~L.~W.\ 2001, \apj, 548, 592 

\bibitem[Shetrone et al.(2003)]{shetrone03} Shetrone, M., Venn, 
K.~A., Tolstoy, E., et al.\ 2003, \aj, 125, 684 

\bibitem[Shen et al.(2015)]{shen15} Shen, S., Cooke, R.~J., 
Ramirez-Ruiz, E., et al.\ 2015, \apj, 807, 115 

\bibitem[Simon et al.(2010)]{simon10} Simon, J.~D., Frebel, A., 
McWilliam, A., Kirby, E.~N., Thompson, I.~B.\ 2010, \apj, 716, 446 

\bibitem[Simon et al.(2015)]{simon15} Simon, J.~D., 
Drlica-Wagner, A., Li, T.~S., et al.\ 2015, \apj, 808, 95 

\bibitem[Siqueira Mello et al.(2013)]{siqueiramello13} 
Siqueira Mello, C., Spite, M., Barbuy, B., et al.\ 2013, \aap, 550, A122 


\bibitem[Smith et al.(2015)]{smith15} Smith, B.~D., Wise, 
J.~H., O'Shea, B.~W., Norman, M.~L., 
Khochfar, S.\ 2015, \mnras, 452, 2822 

\bibitem[Sneden(1973)]{sneden73} Sneden, C.~A.\ 1973,
Ph.D.~Thesis, Univ.\ of Texas at Austin

\bibitem[Sneden et al.(1994)]{sneden94} Sneden, C., Preston, 
G.~W., McWilliam, A., Searle, L.\ 1994, \apjl, 431, L27 

\bibitem[Sneden et al.(1996)]{sneden96} Sneden, C., McWilliam, 
A., Preston, G.~W., et al.\ 1996, \apj, 467, 819 

\bibitem[Sneden et al.(2003)]{sneden03} Sneden, C., Cowan, 
J.~J., Lawler, J.~E., et al.\ 2003, \apj, 591, 936 

\bibitem[Sneden et al.(2008)]{sneden08} Sneden, C., Cowan, J.~J., 
Gallino, R.\ 2008, \araa, 46, 241 

\bibitem[Sneden et al.(2009)]{sneden09} Sneden, C., Lawler, J.~E., 
Cowan, J.~J., Ivans, I.~I., Den Hartog, E.~A.\ 2009, \apjs, 182, 80 

\bibitem[Sobeck et al.(2007)]{sobeck07} Sobeck, J.~S., Lawler, 
J.~E., Sneden, C.\ 2007, \apj, 667, 1267 

\bibitem[Sobeck et al.(2011)]{sobeck11} Sobeck, J.~S., Kraft, 
R.~P., Sneden, C., et al.\ 2011, \aj, 141, 175 

\bibitem[Starkenburg et al.(2014)]{starkenburg14} Starkenburg, E., 
Shetrone, M.~D., McConnachie, A.~W., 
Venn, K.~A.\ 2014, \mnras, 441, 1217 

\bibitem[Tafelmeyer et al.(2010)]{tafelmeyer10} Tafelmeyer, M., 
Jablonka, P., Hill, V., et al.\ 2010, \aap, 524, A58 

\bibitem[Truran et al.(2002)]{truran02} Truran, J.~W., Cowan, 
J.~J., Pilachowski, C.~A., Sneden, C.\ 2002, \pasp, 114, 1293 

\bibitem[Tsujimoto \& Shigeyama(2014)]{tsujimoto14} Tsujimoto, T., 
Shigeyama, T.\ 2014, \aap, 565, L5 

\bibitem[Tsujimoto et al.(2015)]{tsujimoto15} Tsujimoto, T., 
Ishigaki, M.~N., Shigeyama, T., Aoki, W.\ 2015, \pasj, 67, L3 

\bibitem[Udry et al.(1999)]{udry99} Udry, S., Mayor, M., 
Maurice, E., et al.\ 1999, ASP Conf.\ Ser.:\ Precise Stellar Radial 
Velocities, 185, 383 

\bibitem[Umeda \& Nomoto(2005)]{umeda05} Umeda, H., 
Nomoto, K.\ 2005, \apj, 619, 427 

\bibitem[van de Voort et al.(2015)]{vandevoort15} van de Voort, F., 
Quataert, E., Hopkins, P.~F., Kere{\v s}, D., 
Faucher-Gigu{\`e}re, C.-A.\ 2015, \mnras, 447, 140 

\bibitem[Venn et al.(2012)]{venn12} Venn, K.~A., Shetrone, 
M.~D., Irwin, M.~J., et al.\ 2012, \apj, 751, 102 

\bibitem[Walker et al.(2015)]{walker15} Walker, M.~G., Mateo, 
M., Olszewski, E.~W., et al.\ 2015, \apj, 808, 108 

\bibitem[Wallner et al.(2015)]{wallner15} Wallner, A., 
Faestermann, T., Feige, J., et al.\ 2015, Nature Communications, 6, 5956 

\bibitem[Wanajo et al.(2003)]{wanajo03} Wanajo, S., Tamamura, 
M., Itoh, N., et al.\ 2003, \apj, 593, 968 

\bibitem[Webster et al.(2015)]{webster15} Webster, D., Frebel, A., 
Bland-Hawthorn, J.\ 2015, \apj, submitted (arXiv:1509.00856)

\bibitem[Westmeier et al.(2015)]{westmeier15} Westmeier, T., 
Staveley-Smith, L., Calabretta, M., et al.\ 2015, \mnras, 453, 338 

\bibitem[Wickliffe et al.(1994)]{wickliffe94} Wickliffe, M.~E., 
Salih, S., Lawler, J.~E.\ 1994, \jqsrt, 51, 545 

\bibitem[Wickliffe et al.(2000)]{wickliffe00} Wickliffe, M.~E., 
Lawler, J.~E., Nave, G.\ 2000, \jqsrt, 66, 363 

\bibitem[Wood et al.(2013)]{wood13} Wood, M.~P., Lawler, J.~E., 
Sneden, C., Cowan, J.~J.\ 2013, \apjs, 208, 27

\bibitem[Wood et al.(2014)]{wood14} Wood, M.~P., Lawler, 
J.~E., Sneden, C., Cowan, J.~J.\ 2014, \apjs, 211, 20 

\bibitem[Woosley \& Hoffman(1992)]{woosley92} Woosley, S.~E., 
Hoffman, R.~D.\ 1992, \apj, 395, 202 




\end{thebibliography}
\end{document}